\documentclass[screen,nonacm]{acmart}

\AtBeginDocument{%
  }

\setcopyright{acmcopyright}
\copyrightyear{2023}
\acmYear{2023}
\acmDOI{XXXXXXX.XXXXXXX}

\acmPrice{15.00}
\acmISBN{978-1-4503-XXXX-X/18/06}

\newcommand{\bra}[1]{\langle {#1} |}
\newcommand{\ket}[1]{| {#1} \rangle}

\newcommand{\ketbra}[1]{| {#1} \rangle\langle {#1} |}

\newcommand{\lpnorm}[2]{\left\|#2\right\|_{#1}}
\newcommand{\trdist}[1]{\left\|#1\right\|_{\text{tr}}}
\newcommand{\fidelity}[2]{F\left(#1,#2\right)}
\newcommand{\tr}[1]{\text{tr}\left[#1\right]}
\newcommand{\ptr}[2]{\text{tr}_{#1}\left[#2\right]}
\newcommand{\diamondnorm}[1]{\left\|#1\right\|_{\diamond}}

\newcommand{\linop}[1]{\mathbf{L}\left(#1\right)}
\newcommand{\pos}[1]{\mathbf{Pos}\left(#1\right)}

\newcommand{\dop}[1]{\mathbf{S}\left(#1\right)}
\newcommand{\puredop}[1]{\mathbf{P}\left(#1\right)}
\newcommand{\proj}[1]{\mathbf{Proj}\left(#1\right)}
\newcommand{\idop}{\mathbb{I}}

\newcommand{\hh}{\mathcal{H}}
\newcommand{\kk}{\mathcal{K}}
\newcommand{\cd}{\mathbb{C}^d}
\newcommand{\cdim}[1]{\mathbb{C}^{#1}}
\newcommand{\rdim}[1]{\mathbb{R}^{#1}}
\newcommand{\vspan}[1]{\text{span}\left(#1\right)}
\newcommand{\cc}{\mathbb{C}}
\newcommand{\rr}{\mathbb{R}}

\newcommand{\re}[1]{\text{Re}\left[#1\right]}

\newcommand{\polylog}[1]{polylog\left(#1\right)}
\newcommand{\poly}[1]{poly\left(#1\right)}

\newcommand{\conv}[1]{\text{conv}\left(#1\right)}

\usepackage[normalem]{ulem}

\begin{document}

\title{Probabilistic unitary synthesis with optimal accuracy}

\author{Seiseki Akibue}
\email{seiseki.akibue@ntt.com}
\orcid{0000-0001-9654-9361}
\affiliation{%
  \institution{NTT Communication Science Laboratories, NTT Corporation}
  \streetaddress{3--1, Morinosato-Wakamiya}
  \city{Atsugi}
  \state{Kanagawa}
  \country{Japan}
  \postcode{243-0198}
}

\author{Go Kato}
\affiliation{%
  \institution{Advanced ICT Research Institute, NICT}
  \streetaddress{4--2--1, Nukui-Kitamachi}
  \city{Koganei}
  \state{Tokyo}
  \country{Japan}
  \postcode{184-8795}
}

\author{Seiichiro Tani}
\affiliation{%
  \institution{Department of Mathematics, Waseda University}
  \streetaddress{1--6--1, Nishi-Waseda}
  \city{Shinjuku}
  \state{Tokyo}
  \country{Japan}
  \postcode{169-8050}
}


\begin{abstract}
The purpose of unitary synthesis is to find a gate sequence that optimally approximates a target unitary transformation. A new synthesis approach, called probabilistic synthesis, has been introduced, and its superiority has been demonstrated over traditional deterministic approaches with respect to approximation error and gate length. 
However, the optimality of current probabilistic synthesis algorithms is unknown.
We obtain the tight lower bound on the approximation error obtained by the optimal probabilistic synthesis, which guarantees the sub-optimality of current algorithms. We also show its tight upper bound, which improves and unifies current upper bounds depending on the class of target unitaries. These two bounds reveal the fundamental relationship of approximation error between probabilistic approximation and deterministic approximation of unitary transformations.
From a computational point of view, we show that the optimal probability distribution can be computed by the semidefinite program (SDP) we construct. We also construct an efficient probabilistic synthesis algorithm for single-qubit unitaries, rigorously estimate its time complexity, and show that it reduces the approximation error quadratically compared with deterministic algorithms.
\end{abstract}



\received{16 January 2023}
\received[revised]{16 March 2009}
\received[accepted]{5 June 2009}

\maketitle

\section{Introduction}
In quantum simulation and quantum computation, a global unitary transformation on a many-body quantum system is obtained as a sequence of unitary transformations on a fixed-size system, e.g., those obtained by nearest-neighbor interactions. To guarantee and increase the accuracy of obtaining such transformations, rather than controlling their continuous parameters, each unitary transformation on the fixed-size system is realized as a sequence of gates chosen from a finite {\it gate set} $\{g_i\}_{i}$, where each $g_i$ results in a fixed unitary transformation with negligible error thanks to the sophisticated calibration, quantum error correction \cite{B15} or the nature of the system \cite{K03}.
If $\{g_i\}_{i}$ is {\it universal}, arbitrary unitary transformation can be approximated by a unitary transformation $g_{i_n}\circ \cdots\circ g_{i_2}\circ g_{i_1}$ obtained as a gate sequence for an appropriate choice of gate length $n$ depending on the approximate error one wants to achieve. For a given universal gate set such as the set of the Hadamard, controlled-NOT, and $\pi/8$ gates \cite{NCBook},
an algorithm to find a gate sequence for a given unitary transformation and an approximation error bound is called a {\it unitary synthesis} algorithm.


To suppress the effect of decoherence or overhead caused by the fault-tolerant implementation of each gate \cite{ERW98,DM08},
various studies \cite{KBook, ABI02, KMM16, KMM13, R15, BRS15, F11, BG21} have proposed unitary synthesis algorithms for minimizing the length of the output gate sequence. Following the celebrated Solovay-Kitaev algorithm \cite{KBook}, many algorithms are used to find one of the shortest gate sequences that can approximate a target unitary transformation $\Upsilon$ within the desired approximation error. Obviously, the goal can be achieved by brute force search \cite{F11}. However, to guarantee their efficiency, many algorithms are designed for synthesizing restricted classes of unitary transformations by using particular gate sets or for finding a {\it nearly} shortest gate sequence.

While approximating an $\Upsilon$ by using a single sequence of gates is a natural approach, the advantage of another approach using a probabilistic mixture of unitaries has been demonstrated \cite{H17, C17, KLMPP22}.
Suppose that a synthesis algorithm produces a gate sequence for implementing a unitary transformation in $\{\Upsilon_{\vec{i}}\}=\{g_{i_n}\circ \cdots\circ g_{i_2}\circ g_{i_1}\}_{\vec{i}}$ in accordance with the probability distribution $p(\vec{i})$ to approximate an $\Upsilon$.
If the algorithm independently samples $\vec{i}$ for each time the $\Upsilon$ is used in the entire circuit, the physical transformation governed by the randomly executed unitary transformation $\Upsilon_{\vec{i}}$ in accordance with the $p(\vec{i})$ is described by a probabilistic mixture $\sum_{\vec{i}}p(\vec{i})\Upsilon_{\vec{i}}$ of unitaries. In this case, the approximation error should be measured by the distance between the $\Upsilon$ and $\sum_{\vec{i}}p(\vec{i})\Upsilon_{\vec{i}}$. 

Campbell \cite{C17} and Vadym et al. \cite{KLMPP22} constructed algorithms to compute a probability distribution $\{p(\vec{i})\}_{\vec{i}}$ for a given $\Upsilon$ and a set $\{\Upsilon_{\vec{i}}\}_{\vec{i}}$ of unitary transformations implemented as a gate sequence such that the approximation error of $\sum_{\vec{i}}p(\vec{i})\Upsilon_{\vec{i}}$ against $\Upsilon$ is almost quadratically better than that of a single optimal unitary transformation in $\{\Upsilon_{\vec{i}}\}_{\vec{i}}$. 
More precisely, $\diamondnorm{\Upsilon-\sum_{\vec{i}}p(\vec{i})\Upsilon_{\vec{i}}}=O(\epsilon^2)$ for the worst approximation error $\epsilon=\max_\Upsilon\min_{\vec{i}}\frac{1}{2}\diamondnorm{\Upsilon-\Upsilon_{\vec{i}}}$ caused by deterministic synthesis, where $\diamondnorm{\mathcal{A}-\mathcal{B}}$ is the diamond norm \cite{WBook, KBook}.
This also indicates that probabilistically executing $\Upsilon_{\vec{i}}$ in accordance with $p(\vec{i})$ can further reduce the length of the shortest gate sequence without increasing the approximation error (if one measures the error by using the above diamond norm) \cite{C17}.
In general, probabilistic synthesis consists of two procedures; (i) computing a probability distribution $\{p(\vec{i})\}_{\vec{i}}$ and sampling a description $\vec{i}$ of a gate sequence with a classical computer, and (ii) implementing the sampled gate sequence on a quantum computer. In contrast to deterministic synthesis, procedure (ii) may require a quantum computer to rearrange a gate sequence each time it realizes a target unitary transformation. However, such rearrangeability is usually assumed as a standard functionality of a quantum computer.

However, procedure (i) should be meticulously designed to construct a practical synthesis algorithm. This is because the number of possible gate sequences grows exponentially with respect to the length $n$ of the sequence, resulting in a large degree of freedom in choosing $\{p(\vec{i})\}_{\vec{i}}$. 
While a probabilistic synthesis algorithm with guaranteed time complexity exists for single-qubit unitary transformations that correspond to axial rotations \cite{KLMPP22}, no such algorithm was known even for general single-qubit unitary transformations.
Furthermore, a fundamental question remained open regarding the optimality of existing synthesis algorithms in comparison to the minimum approximation error $\min_p\diamondnorm{\Upsilon-\sum_{\vec{i}}p(\vec{i})\Upsilon_{\vec{i}}}$.
Minimax optimization makes it difficult to investigate the minimum approximation error from an analytical perspective except for a few specific $\Upsilon$ and sets $\{\Upsilon_{\vec{i}}\}_{\vec{i}}$ \cite{SS17}.

\subsection{Our contribution}
We obtain the tight lower bound on $\min_p\diamondnorm{\Upsilon-\sum_{\vec{i}}p(\vec{i})\Upsilon_{\vec{i}}}$, which reveals the fundamental limitation of probabilistic synthesis and indicates the sub-optimality of current algorithms. To obtain the main result, we focus on the analytical relationship between $\min_p\diamondnorm{\Upsilon-\sum_{\vec{i}}p(\vec{i})\Upsilon_{\vec{i}}}$ and $\min_{\vec{i}}\diamondnorm{\Upsilon-\Upsilon_{\vec{i}}}$, which represent the minimum approximation error obtained by probabilistic synthesis and that by deterministic synthesis, respectively.
To be mathematically comprehensive, we also obtain the tight upper bound on $\min_p\diamondnorm{\Upsilon-\sum_{\vec{i}}p(\vec{i})\Upsilon_{\vec{i}}}$, which essentially unifies various upper bounds \cite{H17, C17, KLMPP22} depending on the class of target unitary transformations. More precisely, the two bounds are given as the following theorem.

\begin{figure}[h]
  \centering
  \includegraphics[width=.7\linewidth]{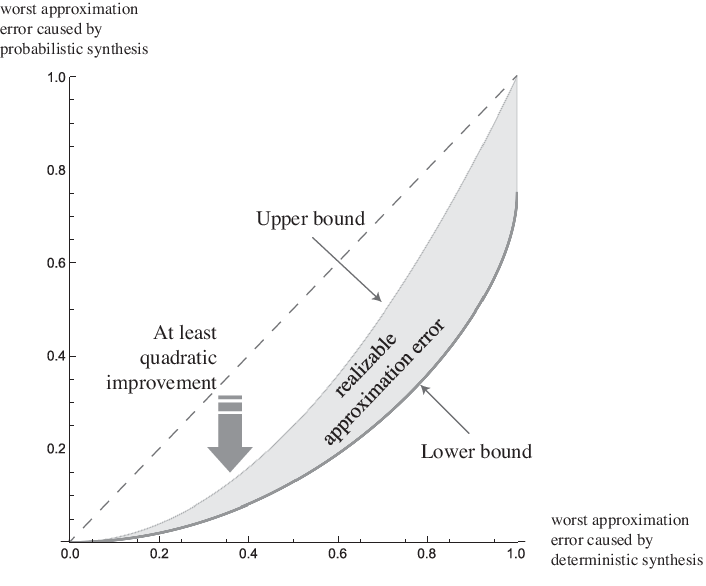}
\caption{\label{fig:accuracybounds}
Lower and upper bounds on worst approximation error $\max_\Upsilon\min_{p}\frac{1}{2}\diamondnorm{\Upsilon-\sum_{\vec{i}}p(\vec{i})\Upsilon_{\vec{i}}}$ caused by probabilistic synthesis with respect to $\max_\Upsilon\min_{\vec{i}}\frac{1}{2}\diamondnorm{\Upsilon-\Upsilon_{\vec{i}}}$ caused by deterministic synthesis for two-qubit systems, i.e., $d=4$.
Both lower and upper bounds, represented with thick and thin curves, respectively, are achievable for certain $\{\Upsilon_{\vec{i}}\}$.}
\end{figure}

THEOREM \ref{thm:main1}. (simplified version)
{\it For an integer $d\geq2$ specified below, let $\Upsilon$ and $\{\Upsilon_{\vec{i}}\}_{\vec{i}}$ be a target unitary transformation and a finite set of unitary transformations on the $d$-dimensional Hilbert space, respectively. It then holds that
\begin{equation}
\label{ineq:worstbound}
 \frac{4\delta}{d}\left(1-\frac{\delta}{d}\right)\leq \max_{\Upsilon}\min_{p}\frac{1}{2}\diamondnorm{\Upsilon-\sum_{\vec{i}}p(\vec{i})\Upsilon_{\vec{i}}}\leq\epsilon^2\ \ {\rm with}\ 
\left\{\begin{array}{l}
  \delta=1-\sqrt{1-\epsilon^2}\ \ \ {\rm and}\\
 \epsilon=\max_\Upsilon\min_{\vec{i}}\frac{1}{2}\diamondnorm{\Upsilon-\Upsilon_{\vec{i}}}.
\end{array}\right.
\end{equation}
}
This theorem provides bounds on the worst approximation error caused when one probabilistically synthesizes the target unitary transformation that is most difficult to approximate.
As shown in Fig. \ref{fig:accuracybounds}, the gap between the upper and lower bounds exists if and only if $d\geq3$. We can show that the gap is inevitable by constructing $\{\Upsilon_{\vec{i}}\}_{\vec{i}}$ for achieving the upper bound and that for achieving the lower bound.
That is, Ineq.~\eqref{ineq:worstbound} represents the fundamental relationship of the approximation error between the deterministic approximation of unitary transformations and their probabilistic approximation that depends only on the dimension $d$ of the system.

From a computational point of view, we show that the optimal probability distribution for approximating an $\Upsilon$ can be computed by the semidefinite program (SDP) we construct when the set $\{\Upsilon_{\vec{i}}\}_{\vec{i}}$ of unitary transformations implemented as a gate sequence is given. (This set is computable with certain synthesis algorithms.)
In addition to its optimality, we can rigorously estimate the worst time complexity of our SDP due to established methods for numerically solving SDPs.
As the second main result, we construct a probabilistic synthesis algorithm for single-qubit unitary transformations from the following theorem.

THEOREM \ref{thm:main2}. (informal version)
{\it For a given gate set, there exists a probabilistic synthesis algorithm for a single-qubit unitary transformation with

{\rm INPUT}: a target single-qubit unitary transformation $\Upsilon$ and target approximation error $\epsilon\in\left(0,1\right)$

{\rm OUTPUT}: a gate sequence implementing a single-qubit unitary transformation $\Upsilon_{\vec{i}}$ sampled from a set $\{\Upsilon_{\vec{i}}\}_{\vec{i}}$ in accordance with probability distribution $\hat{p}(\vec{i})$.

\noindent
such that the algorithm satisfies the following properties:
\begin{itemize}
 \item {\rm Efficiency}: All steps of the algorithm take $\polylog{\frac{1}{\epsilon}}$-time,
 
 \item {\rm Quadratic improvement}: The approximation error $\frac{1}{2}\diamondnorm{\Upsilon-\sum_{\vec{i}}\hat{p}(\vec{i})\Upsilon_{\vec{i}}}$ obtained with this algorithm is upper bounded by $\epsilon^2$, whereas the error $\min_{\vec{i}}\frac{1}{2}\diamondnorm{\Upsilon-\Upsilon_{\vec{i}}}$ obtained by deterministic synthesis using the unitary transformations in $\{\Upsilon_{\vec{i}}\}_{\vec{i}}$ is upper bounded by $\epsilon$.
 
\end{itemize}}

The first property of the algorithm is desirable for fault-tolerant quantum computation (FTQC). The $\polylog{\frac{1}{\epsilon}}$-time overhead due to the synthesis algorithm does not impair a quadratic speedup achieved with a quantum computer over a classical computer since the approximation error of each unitary transformation should satisfy $\frac{1}{\epsilon}=\poly{n}$ if a quantum circuit contains a polynomial number of single-qubit unitaries with respect to the problem size $n$.
Due to the second property of the algorithm, we can verify that it surpasses current algorithms \cite{H17, C17} with respect to the approximation error. 
Our algorithm also surpasses a current algorithm \cite{KLMPP22} in terms of applicability to a general single-qubit unitary transformation.

\subsection{Technical outline}
Previous studies searched for the mixing probability distribution $\{p(\vec{i})\}_{\vec{i}}$ by using the first-order approximation of unitary operators  \cite{H17, C17} and obtained the upper bound on the worst approximation error $\max_{\Upsilon}\min_{p}\frac{1}{2}\diamondnorm{\Upsilon-\sum_{\vec{i}}p(\vec{i})\Upsilon_{\vec{i}}}$ caused by probabilistic synthesis. In contrast, we use the strong duality of SDP, essentially equivalent to the minimax theorem, to obtain tight bounds Ineq.~\eqref{ineq:worstbound} obtained by the optimal mixing probability distribution. A similar technique can be found in the analyses of the optimal convex approximation of quantum states by using a restricted set of states \cite{S17} and that of {\it unital} mappings by using unitary transformations \cite{NDQ12}. While inventing tractable upper bounds on the approximation error of a general unital mapping is an open problem \cite{NDQ12}, we provide an upper bound by exploiting the property of a unitary transformation as a {\it pure} unital mapping.

To prove that our single-qubit unitary synthesis algorithm satisfies the expected properties, we show the fact that $\Upsilon_{\vec{i}}$ that is far from $\Upsilon$ is not necessary to be sampled to optimally approximate $\Upsilon$ for single-qubit unitary transformations by exploiting the {\it magic basis} \cite{BDSW96} representation of single-qubit unitary operators. The magic basis representation enables us to embed the metric space of single-qubit unitary transformations induced by the diamond norm into that of $S^3$ with respect to the angle. While numerical simulations indicate the same fact holds for qudit unitary transformations, rigorous proof is a subject for future work.

\subsection{Organization}
This article is organized as follows. Section \ref{sec:preliminaries} is devoted to preliminaries, introducing basis notations in quantum information theory and semidefinite programming. In Section \ref{sec:SDP}, we construct an SDP that computes the optimal probability distribution in probabilistic synthesis. The SDP is provided as a primal and dual problem whose solutions coincide due to the strong duality of the SDP. The coincidence plays a crucial role in the proof of the first main theorem about the fundamental limitation on the approximation error shown in Section \ref{sec:bounds}. Section \ref{sec:1Q} provides an efficient probabilistic synthesis algorithm for single-qubit unitary transformations as the second main theorem. We also provide a simple geometric interpretation of the superiority of probabilistic synthesis by considering single-qubit unitary transformations corresponding to axial rotations in Section \ref{sec:example1Q}.
We present our conclusions in Section \ref{sec:conclusion}.

\section{Preliminaries}
\label{sec:preliminaries}
In this section, we summarize basic notations used throughout the paper. Note that we consider only finite-dimensional Hilbert spaces. In particular, two-dimensional Hilbert space $\cdim{2}$ is called a qubit.
The $\linop{\hh}$ and $\pos{\hh}$ represent the set of linear operators and positive semidefinite operators on Hilbert space $\hh$, respectively.
$\idop\in\pos{\hh}$ represents the identity operator, and we sometimes use the subscript to specify the system where $\idop$ acts as $\idop_\hh$.
For Hermitian operators $A$ and $B$ on $\hh$, $A\geq B$ represents $A-B\in\pos{\hh}$, and $A>B$ represents $A-B$ is positive definite.
The $\dop{\hh}:=\left\{\rho\in\pos{\hh}:\tr{\rho}=1\right\}$ and $\puredop{\hh}:=\left\{\rho\in\dop{\hh}:\tr{\rho^2}=1\right\}$ represent the set of quantum states and that of pure states, respectively. Pure state $\phi\in\puredop{\hh}$ is sometimes alternatively represented by complex unit vector $\ket{\phi}\in\hh$ satisfying $\phi=\ketbra{\phi}$. Any physical transformation of the quantum state can be represented by a completely positive and trace preserving (CPTP) linear mapping $\Gamma:\linop{\hh_1}\rightarrow\linop{\hh_2}$. There exists one-to-one correspondence between a linear mapping $\Xi:\linop{\hh_1}\rightarrow\linop{\hh_2}$ and its Choi-Jamio\l kowski operator $J(\Xi):=\sum_{i,j}\ket{i}\bra{j}\otimes\Xi(\ket{i}\bra{j})\in\linop{\hh_1\otimes\hh_2}$.
 
The trace distance $\trdist{\rho-\sigma}$ of two quantum states $\rho,\sigma\in\dop{\hh}$ is defined as $\trdist{M}:=\frac{1}{2}\tr{\sqrt{MM^\dag}}$ for $M\in\linop{\hh}$. It represents the maximum total variation distance between probability distributions obtained from measurements performed on two quantum states.
A similar notion measuring the distinguishability of $\rho$ and $\sigma$ is the fidelity function, defined by $\fidelity{\rho}{\sigma}:=\max\tr{\Phi^\rho\Phi^\sigma}$, where $\Phi^\rho\in\puredop{\hh\otimes\hh'}$ is a purification of $\rho$, i.e., $\rho=\ptr{\hh'}{\Phi^\rho}$, and the maximization is taken over all the purifications. Fuchs-van de Graaf inequalities \cite{FG99} provide relationships between the two measures with respect to the distinguishability as follows:
\begin{equation}
\label{ineq:FG}
 1-\sqrt{\fidelity{\rho}{\sigma}}\leq\trdist{\rho-\sigma}\leq\sqrt{1-\fidelity{\rho}{\sigma}}
\end{equation}
holds for any state $\rho,\sigma\in\dop{\hh}$, where the equality of the right inequality holds when $\rho$ and $\sigma$ are pure. 

The distance measuring the distinguishability of two CPTP mappings $\mathcal{A},\mathcal{B}:\linop{\hh_1}\rightarrow\linop{\hh_2}$ corresponding to the trace distance is the diamond norm $\diamondnorm{\mathcal{A}-\mathcal{B}}$ defined by $\frac{1}{2}\diamondnorm{\mathcal{A}-\mathcal{B}}:=\max_{\Phi\in\puredop{\hh_1\otimes\hh_3}}\trdist{((\mathcal{A}-\mathcal{B})\otimes id)(\Phi)}$, where $id$ represents the identity mapping acting on $\hh_3$.

Let $\Xi:\linop{\hh_1}\rightarrow\linop{\hh_2}$ be a linear Hermitian-preserving mapping and $A$ and $B$ be Hermitian operators on $\hh_1$ and $\hh_2$, respectively.
SDP is an optimization problem formally defined with a triple $(\Xi,A,B)$ as follows \cite{WBook}:
\begin{equation}
\begin{tabular}{rlcrl}
\multicolumn{2}{c}{\underline{{\rm Primal problem}}} &\ \ \ \ \ \ \ \ \ \ \ \ \ \  
 &\multicolumn{2}{c}{\underline{{\rm Dual problem}}}\\
{\rm maximize:}&$\tr{AX}$&&{\rm minimize:}&$\tr{BY}$ \\
{\rm subject to:}& $X\in\pos{\hh_1}$,&&
{\rm subject to:}& $Y\text{\ is\ a\ Hermitian\ operator\ on\ }\hh_2$,\\
&$\Xi(X)=B$&&&$\Xi^\dag(Y)\geq A$,
\end{tabular} 
\end{equation}
where $\Xi^\dag:\linop{\hh_2}\rightarrow\linop{\hh_1}$ is the adjoint of $\Xi$, defined as the linear mapping satisfying $\tr{Y^\dag\Xi(X)}=\tr{(\Xi^\dag(Y))^\dag X}$ for all $X\in\linop{\hh_1}$ and $Y\in\linop{\hh_2}$.
We can easily verify that the solution to the primal problem is smaller than or equal to that of the dual problem. The situation when the two solutions coincide is called a {\it strong duality}. Slater's theorem states that the strong duality holds if either of the following conditions holds:
\begin{enumerate}
 \item The solution to the primal problem is finite, and there exists a Hermitian operator $Y$ on $\hh_2$ such that $\Xi^\dag(Y)>A$.
 \item The solution to the dual problem is finite, and there exists a positive definite operator $X$ on $\hh_1$ such that $\Xi(X)=B$.
\end{enumerate}

For a metric space $(X,d)$ and two subsets $S,T\subseteq X$, $S$ is called an $\epsilon$-covering of $T$ if $\sup_{t\in T}\inf_{s\in S}d(s,t)\leq\epsilon$. In this article, we basically assume that $X$ is the set of CPTP mappings, the metric is defined as $d(\mathcal{A},\mathcal{B})=\frac{1}{2}\diamondnorm{\mathcal{A}-\mathcal{B}}$, $S$ is a finite set of unitary transformations and $T$ is a subset of unitary transformations such as a $2\epsilon$-ball $\left\{\Upsilon':\frac{1}{2}\diamondnorm{\Upsilon'-\Upsilon}\leq 2\epsilon\right\}$ around a unitary transformation $\Upsilon:\linop{\hh}\rightarrow\linop{\hh}$.

\section{Semidefinite programming for computing optimal mixing probability}
\label{sec:SDP}
In this section, we construct an SDP for computing the optimal probability distribution that minimizes the diamond norm between the target CPTP mapping $\mathcal{A}$ and a probabilistic mixture of CPTP mappings $\{\mathcal{B}_x\}_{x}$. We can compute the optimal probability distribution in probabilistic unitary synthesis by solving this SDP by restricting $\mathcal{A}$ and $\{\mathcal{B}_x\}_{x}$ as unitary transformations. We also mention the relationship between our SDP and the algorithm proposed by Campbell \cite{C17}.

\begin{proposition}
\label{prop:SDP}
Let $\mathcal{A}$ and $\{\mathcal{B}_x\}_{x\in X}$ be a target CPTP mapping and a finite set of CPTP mappings from $\linop{\hh_1}$ to $\linop{\hh_2}$, respectively. Then, distance $\min_{p}\frac{1}{2}\diamondnorm{\mathcal{A}-\sum_{x\in X}p(x)\mathcal{B}_x}$ and the optimal probability distribution $\{p(x)\}_{x\in X}$, which minimizes the distance, can be computed with the following SDP:
\begin{equation}
\label{eq:SDP}
\begin{tabular}{rlcrl}
\multicolumn{2}{c}{\underline{{\rm Primal problem}}} &\ \ \ \ \ \ \ \ \ \ \ \ \ \  
 &\multicolumn{2}{c}{\underline{{\rm Dual problem}}}\\
{\rm maximize:}&$\tr{J(\mathcal{A})T}-t$&&{\rm minimize:}&$r\in\mathbb{R}$ \\
{\rm subject to:}& $0\leq T\leq\rho\otimes\idop_{\hh_2}$,&&
{\rm subject to:}& $S\geq0\wedge S\geq J\left(\mathcal{A}-\sum_{x\in X}p(x)\mathcal{B}_x\right)$,\\
&$\rho\in\dop{\hh_1}$&&&$r\idop_{\hh_1}\geq\ptr{\hh_2}{S}$,\\
&$\forall x\in X,\tr{J(\mathcal{B}_x)T}\leq t$.&&&$\forall x\in X,p(x)\geq0$,\\
&&&&$\sum_{x\in X}p(x)\leq 1$.
\end{tabular} 
\end{equation}
Note that the strong duality holds in this SDP, i.e., the optimum primal and dual values are equal.
\end{proposition}

\begin{proof}
Recall that for two CPTP mapping $\mathcal{A}$ and $\mathcal{B}$ from $\linop{\hh_1}$ to $\linop{\hh_2}$, $\frac{1}{2}\diamondnorm{\mathcal{A}-\mathcal{B}}$ can be computed by the following SDP:
\begin{center}
 \begin{tabular}{rlcrl}
\multicolumn{2}{c}{\underline{Primal problem}} &\ \ \ \ \ \ \ \ \ \ \ \ \ \ \ \ \ \ \ \ \ \ \ \ \ \ \ \ \ \ \ \  & \multicolumn{2}{c}{\underline{Dual problem}}\\
maximize: & $\tr{J(\mathcal{A}-\mathcal{B})T}$ && minimize:& $r\in\mathbb{R}$\\
subject to: & $0\leq T\leq\rho\otimes\idop_{\hh_2}$, && subject to: & $S\geq0\wedge S\geq J(\mathcal{A}-\mathcal{B})$,\\
& $\rho\in\dop{\hh_1}$. &&& $r\idop_{\hh_1}\geq\ptr{\hh_2}{S}$.\\
\end{tabular}
\end{center}

\noindent The primal problem can be obtained by observing 
\begin{eqnarray}
 \frac{1}{2}\diamondnorm{\mathcal{A}-\mathcal{B}}&=&\max_{\substack{\Phi\in\puredop{\hh_1\otimes\hh_3}\\\Pi\in\mathbf{Proj}(\hh_2\otimes\hh_3)}}\tr{((\mathcal{A}-\mathcal{B})\otimes id)(\Phi)\Pi}\\
 &=&\max_{T\in\mathbf{T}(\hh_1:\hh_2)}\tr{J(\mathcal{A}-\mathcal{B})T},
\end{eqnarray}
where $\Pi$ is a Hermitian projector acting on $\hh_2\otimes\hh_3$, $\mathbf{T}(\hh_1:\hh_2):=\{T\in\pos{\hh_1\otimes\hh_2}:\exists\rho\in\dop{\hh_1},T\leq\rho\otimes\idop\}$ is called the set of measuring strategies \cite{GJ07} or that of quantum testers \cite{GGP09}, and the last equality was shown by Chiribella et al. \cite[Theorem~10]{GGP09}. To be self-contained, we provide a proof for the equality in Appendix \ref{appendix:tester}, with which the equality can be verified by applying Eq.~\eqref{eq:tester_and_network} with fixing $\Xi=\mathcal{A}-\mathcal{B}$. A formal SDP and the verification of the strong duality are provided in Appendix \ref{appendix:SDP}.

By extending the dual problem of this SDP to include the minimization of probability distribution $\{p(x)\}_{x\in X}$, we obtain Eq.~\eqref{eq:SDP}. Note that the last condition $\sum_{x\in X}p(x)\leq 1$ in the dual problem is different from the condition $\sum_{x\in X}p(x)=1$ of a probability distribution; however, the optimum dual value can be achieved under the latter condition.
Again, a formal SDP and the verification of the strong duality are provided in Appendix \ref{appendix:SDP}.
\end{proof}

For a given $\Upsilon:\linop{\hh}\rightarrow\linop{\hh}$ and a given set $\{\Upsilon_x:\linop{\hh}\rightarrow\linop{\hh}\}_{x\in X}$ of unitary transformations implemented as a gate sequence, which forms an $\epsilon$-covering the set of unitary transformations with sufficiently small $\epsilon$, ``the convex hull finding algorithm" proposed by Campbell \cite{C17} can find a probability distribution $\{\tilde{p}(x)\}_{x\in \tilde{X}}$ such that $\sum_{x\in \tilde{X}}\tilde{p}(x)H_x=0$, where $\Upsilon_x(\rho)=\Upsilon(e^{iH_x}\rho e^{-iH_x})$ and $H_x=O(\epsilon)$ for all $x\in\tilde{X}\subseteq X$. Note that $M=O(\epsilon)$ represents $\lpnorm{\infty}{M}=O(\epsilon)$ as $\epsilon\rightarrow 0$ for a linear operator $M\in\linop{\hh}$ depending on $\epsilon$. By using the dual problem in Proposition \ref{prop:SDP}, we can verify that the distance $\epsilon$, which is achievable by a deterministic unitary synthesis finding the closest $\Upsilon_x$ to approximate $\Upsilon$, can be improved into $O(\epsilon^2)$ by mixing unitaries in accordance with the probability distribution $\{\tilde{p}(x)\}_{x\in \tilde{X}}$ as follows. First, by using the dual problem of the SDP to compute the diamond norm between two CPTP mappings, we obtain
\begin{eqnarray}
 \frac{1}{2}\diamondnorm{\Upsilon-\sum_{x\in \tilde{X}}\tilde{p}(x)\Upsilon_x}=\frac{1}{2}\diamondnorm{id-\sum_{x\in \tilde{X}}\tilde{p}(x)\Upsilon^{-1}\circ\Upsilon_x}\leq\lpnorm{\infty}{\ptr{\hh'}{S}}\\
 \label{ineq:dualcond}
 {\rm with}\  S\geq0\wedge S\geq J(id)-\sum_{x\in \tilde{X}}\tilde{p}(x)J(\Upsilon^{-1}\circ\Upsilon_x),
\end{eqnarray}
where $\hh'$ represents the the output system of $\Upsilon$, which is isomorphic to $\hh$.
Second, by using the Taylor expansions $e^{iH_x}=\idop+iH_x+R_x$, where $R_x=O(\epsilon^2)$, we obtain
\begin{eqnarray}
J(id)-\sum_{x\in \tilde{X}}\tilde{p}(x)J(\Upsilon^{-1}\circ\Upsilon_x)&=&\sum_{x\in \tilde{X}}\tilde{p}(x)\left\{-(R_xJ(id)+J(id) R_x^\dag)-i( H_xJ(id) R_x^\dag- R_xJ(id) H_x) \right\}-P\\
&\leq&\sum_{x\in \tilde{X}}\tilde{p}(x)\Big\{\left(\frac{1}{\lpnorm{\infty}{R_x}}R_xJ(id)R_x^\dag+\lpnorm{\infty}{R_x}J(id)\right)\nonumber\\
\label{ineq:upbound}
&&\ \ \ \ \ \ \ \ \ \ \ \ \ \ \ \ +\left(\frac{\lpnorm{\infty}{R_x}}{\lpnorm{\infty}{H_x}}H_xJ(id)H_x+\frac{\lpnorm{\infty}{H_x}}{\lpnorm{\infty}{R_x}}R_xJ(id)R_x^\dag\right)\Big\}
\end{eqnarray}
where $P=\sum_{x\in\tilde{X}}\tilde{p}(x)(H_xJ(id)H_x+R_xJ(id)R_x^\dag)\in\pos{\hh\otimes\hh'}$, $H_x$ and $R_x$ acts on $\hh'$ and we use the fact that $\ket{\tilde{\phi}}\bra{\tilde{\psi}}+\ket{\tilde{\psi}}\bra{\tilde{\phi}}\leq\tilde{\phi}+\tilde{\psi}$ with complex vectors 
$(\ket{\tilde{\phi}},\ket{\tilde{\psi}})=\left(\lpnorm{\infty}{R_x}^{-\frac{1}{2}}\sum_j(\idop_\hh\otimes R_x)\ket{jj},\lpnorm{\infty}{R_x}^{\frac{1}{2}}\sum_j\ket{jj}\right)$ and 
$(\ket{\tilde{\phi}},\ket{\tilde{\psi}})=\left(\lpnorm{\infty}{R_x}^{\frac{1}{2}}\lpnorm{\infty}{H_x}^{-\frac{1}{2}}\sum_j(\idop_\hh\otimes H_x)\ket{jj},i\lpnorm{\infty}{R_x}^{-\frac{1}{2}}\lpnorm{\infty}{H_x}^{\frac{1}{2}}\sum_j(\idop_\hh\otimes R_x)\ket{jj}\right)$
 in the inequality. Third, by letting $S$ in Eq.~\eqref{ineq:dualcond} be R.H.S. of Eq.~\eqref{ineq:upbound}, we obtain
\begin{eqnarray}
 \lpnorm{\infty}{\ptr{\hh'}{S}}&=&\lpnorm{\infty}{\sum_{x\in \tilde{X}}\tilde{p}(x)\left(\frac{(R_x^\dag R_x)^T}{\lpnorm{\infty}{R_x}}+\lpnorm{\infty}{R_x}\idop_{\hh}+\frac{\lpnorm{\infty}{R_x}(H_x^2)^T}{\lpnorm{\infty}{H_x}}+\frac{\lpnorm{\infty}{H_x}(R_x^\dag R_x)^T}{\lpnorm{\infty}{R_x}}\right)}=O(\epsilon^2).
\end{eqnarray}

Since the approximation error $ \frac{1}{2}\diamondnorm{\Upsilon-\sum_{x\in \tilde{X}}\tilde{p}(x)\Upsilon_x}$ is generally worse than the optimal one  $\min_{p}\frac{1}{2}\diamondnorm{\Upsilon-\sum_{x\in X}p(x)\Upsilon_x}$, we can obtain a better probability distribution and better estimation of the approximation error by numerically solving the SDP shown in Proposition \ref{prop:SDP}. The ellipsoid method guarantees that $\{p(x)\}_{x\in X}$ and $r$ in the dual problem such that the difference between $r$ and the optimum dual value is less than $\epsilon$ can be computed in $\poly{|X|\log\left(\frac{1}{\epsilon}\right)}$-time \cite{L03}. Note that we assume the dimension of the Hilbert space is constant since the unitary synthesis is usually executed for $\Upsilon$ on a fixed-size system.

\section{Tight bounds on error of probabilistic approximation}
\label{sec:bounds}
This section investigates the relationship between the discrete approximation of unitary transformations and the probabilistic approximation for a general $\Upsilon$ and general set $\{\Upsilon_x\}_x$.
Specifically, we show the tight relationship between $\min_p\diamondnorm{\Upsilon-\sum_{x}p(x)\Upsilon_x}$ and $\min_{x}\diamondnorm{\Upsilon-\Upsilon_x}$, where the former represents the minimum approximation error obtained by probabilistic synthesis and the latter represents that by deterministic synthesis when $\{\Upsilon_x\}_x$ is a set of unitary transformations implemented as a gate sequence.
The first lemma shows the fundamental limitation of probabilistic synthesis, and the second one shows its superiority over deterministic synthesis.

\begin{lemma}
\label{lemma:lowerbound}
For an integer $d\geq2$ specified below, let $\Upsilon:\linop{\cd}\rightarrow\linop{\cd}$ and $\left\{\Upsilon_x:\linop{\cd}\rightarrow\linop{\cd}\right\}_{x\in X}$ be a target unitary transformation and finite set of unitary transformations, respectively. Then
\begin{eqnarray}
\label{ineq:lowerbound}
\frac{2}{d}\epsilon^2\leq \frac{4\delta}{d}\left(1-\frac{\delta}{d}\right)\leq \min_{p}\frac{1}{2}\diamondnorm{\Upsilon-\sum_{x\in X}p(x)\Upsilon_x}\ 
{\rm with}\  
\left\{\begin{array}{l}
  \delta=1-\sqrt{1-\epsilon^2}\ \ \ {\rm and}\\
 \epsilon=\min_{x\in X}\frac{1}{2}\diamondnorm{\Upsilon-\Upsilon_x}
\end{array}\right.
 \end{eqnarray}
 holds, where the minimization of $p$ is taken over probability distributions over $X$.
\end{lemma}
\begin{proof}
 The first inequality can be straightforwardly verified as follows:
\begin{equation}
 \frac{2}{d}\epsilon^2=\frac{4\delta}{d}\left(1-\frac{\delta}{2}\right)\leq \frac{4\delta}{d}\left(1-\frac{\delta}{d}\right).
\end{equation}
Thus, we prove the second inequality. First, by computing the diamond norm between $\Upsilon$ and $\Upsilon_x$, we obtain
\begin{eqnarray}
\frac{1}{2}\diamondnorm{\Upsilon-\Upsilon_x}&=&\max_{\Phi\in\puredop{\cd\otimes\cd}}\trdist{\Upsilon\otimes id_{\cd}(\Phi)-\Upsilon_x\otimes id_{\cd}(\Phi)}\\
&=&\max_{\Phi\in\puredop{\cd\otimes\cd}}\sqrt{1-\fidelity{\Upsilon\otimes id_{\cd}(\Phi)}{\Upsilon_x\otimes id_{\cd}(\Phi)}}\\
&=&\sqrt{1-\min_{\Phi\in\puredop{\cd\otimes\cd}}|\bra{\Phi}U^\dag U_x\otimes\idop_{\cd}\ket{\Phi}|^2}\\
\label{eq:L0}
&=&\sqrt{1-\min_{\rho\in\dop{\cd}}|\tr{\rho U^\dag U_x}|^2},
\end{eqnarray}
where $\Upsilon(\rho)=U\rho U^\dag$ and $\Upsilon_x(\rho)=U_x\rho U_x^\dag$. This indicates
\begin{equation}
\label{eq:L1}
 1-\delta=\max_{x\in X}\min_{\rho\in\dop{\cd}}|\tr{\rho U^\dag U_x}|.
\end{equation}

Next, by using the primal problem in our SDP in Proposition \ref{prop:SDP}, we obtain
\begin{eqnarray}
\min_p \frac{1}{2}\diamondnorm{\Upsilon-\sum_{x\in X}p(x)\Upsilon_x}&=&\max_{T\in\mathbf{T}(\cd:\cd)}\left(\tr{J(\Upsilon)T}-\max_{x\in X}\tr{J(\Upsilon_x)T}\right)\\
&\geq&\frac{1}{d^2}\left(\tr{J(\Upsilon)J(\Upsilon)}-\max_{x\in X}\tr{J(\Upsilon_x)J(\Upsilon)}\right)\\
&=&1-\frac{1}{d^2}\max_{x\in X}\left|\tr{U^\dag U_x}\right|^2,
\label{ineq:lbR1}
\end{eqnarray}
where $\mathbf{T}(\hh_1:\hh_2):=\{T\in\pos{\hh_1\otimes\hh_2}:\exists\rho\in\dop{\hh_1},T\leq\rho\otimes\idop_{\hh_2}\}$, and we set $T=\frac{1}{d^2}J(\Upsilon)\left(\leq\frac{\idop_{\cd}}{d}\otimes\idop_{\cd}\right)$ to obtain the inequality.

In Eq.~\eqref{eq:L1} and Eq.~\eqref{ineq:lbR1}, the same unitary operator $W=U^\dag U_x$ appears in the term $\min_{\rho\in\dop{\cd}}|\tr{\rho W}|$ and $|\tr{W}|$, respectively. We can prove the second inequality in Ineq.~\eqref{ineq:lowerbound} by establishing a relationship between the two terms as follows.
For any unitary operator $W$ on $\cd$ $(d\geq2)$,
\begin{eqnarray}
 \frac{1}{d}\left|\tr{W}\right|=\frac{1}{d}\left|\sum_{i=1}^d\lambda_i(W)\right|
 \leq\frac{2}{d}\min_p\left|\sum_{i=1}^dp(i)\lambda_i(W)\right|+\frac{d-2}{d}
=\frac{2}{d}\min_{\rho\in\dop{\cd}}\left|\tr{\rho W}\right|+\frac{d-2}{d}
\end{eqnarray}
holds, where $\lambda_i(W)$ is the $i$-th eigenvalue of $W$, and in the inequality, we use the following two facts: (i) the minimization is achieved only if $p$ satisfies $\forall i,p(i)\leq\frac{1}{2}$ due to a geometric observation, and (ii) for such $p$ and complex numbers $\lambda_i\in\{z\in\mathbb{C}:|z|=1\}$, $\left|\sum_ip(i)\lambda_i\right|\geq\left|\sum_i\frac{1}{2}\lambda_i\right|-\left|\sum_i\left(\frac{1}{2}-p(i)\right)\lambda_i\right|\geq\frac{1}{2}\left|\sum_i\lambda_i\right|-\sum_i\left(\frac{1}{2}-p(i)\right)=\frac{1}{2}\left|\sum_i\lambda_i\right|-\frac{d-2}{2}$. 
\end{proof}

To the best of our knowledge, the dependence of the approximation error obtained by probabilistic synthesis on the dimension of the Hilbert space shown in this theorem has never been found. This dependence is inevitable since we can also show the sharpness of this theorem in Appendix \ref{appendix:sharp}. More precisely, we can show that for any real number $\epsilon\in(0,1]$, any integer $d\geq2$ and any $\Upsilon$, there exists $\{\Upsilon_x\}_{x\in X}$ achieving the lower bound in Ineq.~\eqref{ineq:lowerbound}.

In the following lemma, we show the tight upper bound showing that the worst approximation error caused by deterministic synthesis can be reduced by probabilistic synthesis at least quadratically. Our upper bound slightly improves the various existing upper bounds \cite{H17, C17}, which have been proven for several classes of target unitary transformations and $\epsilon$-coverings $\{\Upsilon_x\}_{x\in X}$ with small $\epsilon$. Using Proposition \ref{prop:axialU}, shown in the next section, we can verify that our upper bound is still tight even if we consider the approximation of axial single-qubit unitary transformations.

\begin{lemma}
\label{lemma:upperbound}
For a non-negative real number $\epsilon\geq0$ and integer $d\geq2$ specified below, if $\left\{\Upsilon_x:\linop{\cd}\rightarrow\linop{\cd}\right\}_{x\in X}$ is a finite $\epsilon$-covering of the set of unitary transformations, i.e., $\max_\Upsilon\min_{x\in X}\frac{1}{2}\diamondnorm{\Upsilon-\Upsilon_x}\leq\epsilon$, then
\begin{equation}
\label{ineq:upperbound}
\min_{p}\frac{1}{2}\diamondnorm{\Upsilon-\sum_{x\in X}p(x)\Upsilon_x}\leq\epsilon^2
\end{equation}
holds for any unitary transformation $\Upsilon$, where the minimization of $p$ are taken over probability distributions over $X$.
\end{lemma}
\begin{proof}
First, by using the primal problem in our SDP in Proposition \ref{prop:SDP}, we obtain
\begin{eqnarray}
(L.H.S.)
&=&\max_{T\in\mathbf{T}(\cd:\cd)}\left(\tr{J(\Upsilon)T}-\max_{x\in X}\tr{J(\Upsilon_x)T}\right)\\
\label{eq:R2}
&=&\max_{\substack{\Phi\in\puredop{\cd\otimes\hh}\\\Pi\in\mathbf{Proj}(\cd\otimes\hh)}}
  \Big(\tr{(U\otimes\idop_{\hh})\Phi (U\otimes\idop_{\hh})^\dag\Pi}
  -\max_{x\in X}\tr{(U_x\otimes\idop_{\hh})\Phi (U_x\otimes\idop_{\hh})^\dag\Pi}\Big),
\label{eq:R1}
\end{eqnarray}
where $\mathbf{T}(\hh_1:\hh_2):=\{T\in\pos{\hh_1\otimes\hh_2}:\exists\rho\in\dop{\hh_1},T\leq\rho\otimes\idop_{\hh_2}\}$, $\Upsilon(\rho)=U\rho U^\dag$, $\Upsilon_x(\rho)=U_x\rho U_x^\dag$, $\mathbf{Proj}(\hh)$ is the set of Hermitian projectors on $\hh$, and we use Eq.~\eqref{eq:tester_and_network} by taking $\Xi\in\{\Upsilon-\Upsilon_x\}_{x\in X}$ to obtain the last equality.

Let $\hat{\Phi}$ and $\hat{\Pi}$ maximize Eq.~\eqref{eq:R2}. We can verify that $\hat{\Pi}U\ket{\hat{\Phi}}=0$ if and only if there exists $x\in X$ such that $\Upsilon_x=\Upsilon$. 
If $\hat{\Pi}U\ket{\hat{\Phi}}\neq0$, let $\hat{\Psi}$ be the pure state such that $\ket{\hat{\Psi}}\propto\hat{\Pi}U\ket{\hat{\Phi}}$. Then, we can verify that Eq.~\eqref{eq:R2} is still maximized even if we replace $\hat{\Pi}$ with $\hat{\Psi}$.
If $\hat{\Pi}U\ket{\hat{\Phi}}=0$, $(\exists x\in X,\Upsilon_x=\Upsilon)$ indicates that Eq.~\eqref{eq:R2} is still maximized even if we replace $\hat{\Pi}$ and $\hat{\Phi}$ with an arbitrary pure state $\hat{\Psi}$ and $(\Upsilon^{-1}\otimes id_\hh)(\hat{\Psi})$, respectively.
Thus, in both cases, $\Pi$ in Eq.~\eqref{eq:R2} can be restricted as a pure state, i.e., $\Pi=\Psi\in\puredop{\cd\otimes\hh}$, and we proceed as follows:
\begin{eqnarray}
\label{eq:R3}
 {\rm Eq}.~\eqref{eq:R2} =\max_{\Phi,\Psi\in\puredop{\cd\otimes\hh}}\Big(|\bra{\Psi}U\otimes\idop_{\hh}\ket{\Phi}|^2 -\max_{x\in X}|\bra{\Psi}U_x\otimes\idop_{\hh}\ket{\Phi}|^2\Big).
  \end{eqnarray}
Before proceeding to the next step, we show that the set of mappings $f_{\Phi,\Psi}:U\mapsto |\bra{\Psi}U\otimes\idop_{\hh}\ket{\Phi}|$ associated with pure states $\Phi$ and $\Psi$ is equivalent to that of mappings $g_A:U\mapsto\left|\tr{A U}\right|$ associated with linear operator $A\in\linop{\cd}$ such that $\lpnorm{1}{A}\leq1$, where $\lpnorm{1}{A}$ is the Schatten $1$-norm of $A$. By using decompositions $\ket{\Phi}=\sum_{i,j}\alpha_{ij}\ket{i}\ket{j}$ and $\ket{\Psi}=\sum_{i,j}\beta_{ij}\ket{i}\ket{j}$ with respect to orthonormal bases, we can verify that $g_A$ with $A=\sum_{i,j,k}\alpha_{ik}\beta^*_{jk}\ket{i}\bra{j}$ is equal to $f_{\Phi,\Psi}$ and $\lpnorm{1}{A}=\max_Ug_A(U)=\max_Uf_{\Phi,\Psi}(U)\leq1$. On the other hand, by using the singular value decomposition $A=\sum_ip_i\ket{x_i}\bra{y_i}$, where $\lpnorm{1}{A}\leq1$ indicates $p+\sum_ip_i=1$ with some $p\geq0$, we can verify that $f_{\Phi,\Psi}$ with $\ket{\Phi}=\sqrt{p}\ket{0}\ket{\bot}+\sum_i\sqrt{p_i}\ket{x_i}\ket{i}$ and $\ket{\Psi}=\sqrt{p}\ket{0}\ket{\bot'}+\sum_i\sqrt{p_i}\ket{y_i}\ket{i}$ ($\{\ket{i}\}_i\cup\{\ket{\bot},\ket{\bot'}\}$ is an orthonormal basis) is equal to $g_A$.
  
By using the equivalent between two sets of mappings, we proceed as follows:
\begin{eqnarray}
\label{eq:R4}
 {\rm Eq}.~\eqref{eq:R3}&=&\max_{A:\lpnorm{1}{A}\leq1}\left(\left|\tr{A U}\right|^2-\max_{x\in X}\left|\tr{A U_x}\right|^2\right)
 =\max_{V,\rho\in\dop{\cd}}\left(\left|\tr{\rho V^\dag U}\right|^2-\max_{x\in X}\left|\tr{\rho V^\dag U_x}\right|^2\right),
\end{eqnarray}
where we use the fact that the maximization is achieved when $\lpnorm{1}{A}=1$ and use the polar decomposition $A=\rho V^\dag$ with a unitary operator $V$ acting on $\cd$. 

 By using Eq.~\eqref{eq:R4}, we obtain
\begin{eqnarray}
  \max_{\Upsilon}\min_{p}\frac{1}{2}\diamondnorm{\Upsilon-\sum_{x\in X}p(x)\Upsilon_x}&=&
  \max_{V,\rho\in\dop{\cd}}\left(\max_U\left|\tr{\rho V^\dag U}\right|^2-\max_{x\in X}\left|\tr{\rho V^\dag U_x}\right|^2\right)\\
  \label{eq:worstaccuracy}
  &=&1-\min_{V,\rho\in\dop{\cd}}\max_{x\in X}\left|\tr{\rho V^\dag U_x}\right|^2\\
  &\leq&1-\min_V\max_{x\in X}\min_{\rho\in\dop{\cd}}\left|\tr{\rho V^\dag U_x}\right|^2,
\end{eqnarray}
where the maximization of $\Upsilon$ is taken over unitary transformations, and we use the fact that $\max_x\min_yf(x,y)\leq\min_y\max_xf(x,y)$ for any $f$ if the maximum and minimum exist in the inequality. Using Eq.~\eqref{eq:L0} completes the proof.
\end{proof}

The combination of Lemmas \ref{lemma:lowerbound} and \ref{lemma:upperbound} can be summarized as the following theorem.
\begin{theorem}
 \label{thm:main1}
 For an integer $d\geq2$ specified below, let $\Upsilon:\linop{\cd}\rightarrow\linop{\cd}$ and $\left\{\Upsilon_x:\linop{\cd}\rightarrow\linop{\cd}\right\}_{x\in X}$ be a target unitary transformation and finite set of unitary transformations, respectively. Then,
\begin{eqnarray}
\label{ineq:fundamentalbound}
\frac{4\delta_\Upsilon}{d}\left(1-\frac{\delta_\Upsilon}{d}\right)\leq \min_{p}\frac{1}{2}\diamondnorm{\Upsilon-\sum_{x\in X}p(x)\Upsilon_x}\leq\epsilon^2\ 
{\rm with}\  
\left\{\begin{array}{l}
  \delta_\Upsilon=1-\sqrt{1-\epsilon_\Upsilon^2}\\
 \epsilon_\Upsilon=\min_{x\in X}\frac{1}{2}\diamondnorm{\Upsilon-\Upsilon_x}\\
 \epsilon=\max_{\Upsilon}\min_{x\in X}\frac{1}{2}\diamondnorm{\Upsilon-\Upsilon_x}
\end{array}\right.
 \end{eqnarray}
 holds, where the maximization of $\Upsilon$ and minimization of $p$ are taken over unitary transformations on $\linop{\cd}$ and probability distributions over $X$, respectively.
\end{theorem}
By maximizing $\Upsilon$ over all the unitary transformations, we obtain Ineq.~\eqref{ineq:worstbound} as a simplified version of this theorem.
As mentioned in the introduction, in Appendix \ref{appendix:sharp}, we show that both the upper and lower bounds in Ineq.~\eqref{ineq:worstbound} are tight, i.e., for any real number $\epsilon\in(0,1]$ and any integer $d\geq2$, the two bounds are achievable for some $\{\Upsilon_{\vec{i}}\}_{\vec{i}}$.

\section{Probabilistic synthesis for single-qubit unitary transformation}
\label{sec:1Q}
In this section, we construct a simplified SDP that computes the optimal mixing probability for single-qubit-unitary synthesis. Before discussing that, we first show the special properties of the probabilistic mixture of single-qubit unitaries. In the first subsection, we prove Lemma \ref{lemma:support}, which is a crucial ingredient for constructing the SDP and has a direct application to constructing an efficient probabilistic synthesis algorithm. In the second subsection, we investigate the approximation of single-qubit unitary transformations corresponding to axial rotations to provide a geometric interpretation of the quadratic improvement owing to the probabilistic mixture and confirmation of Lemma \ref{lemma:support}.

We show the first special property of a single-qubit unitary operator in the following Lemma, which essentially shows the equivalence between the set of maximally entangled two-qubit states and a real subspace in the two qubits.

\begin{lemma}
 \label{lemma:1QMES}
 For any finite set $\{\Phi_x\in\puredop{\cdim{2}\otimes\cdim{2}}\}_{x\in X}$ of maximally entangled states and any real numbers $\{r_x\in\rr\}_{x\in X}$,  the Hermitian operator $H=\sum_{x\in X}r_x\Phi_x$ is diagonalizable with respect to maximally entangled eigenstates.
\end{lemma}
\begin{proof}
First, we show the equivalence between the set of two-qubit maximally entangled vectors and a real subspace in the two qubits. Define four vectors representing maximally entangled states:
\begin{eqnarray}
\label{eq:MESbasis}
\ket{\Psi_1}=\frac{1}{\sqrt{2}}(\ket{00}+\ket{11}),\ \ \ &&\ket{\Psi_2}=\frac{i}{\sqrt{2}}(\ket{00}-\ket{11}),\nonumber\\
\ket{\Psi_3}=\frac{i}{\sqrt{2}}(\ket{01}+\ket{10}),\ \ \ &&\ket{\Psi_4}=\frac{1}{\sqrt{2}}(\ket{01}-\ket{10}).
\end{eqnarray}
Any vector in the real subspace $\kk_{MES}$ spanned by $\{\ket{\Psi_i}\}_{i=1}^4$ can be represented by
\begin{equation}
 \label{eq:rsubspace}
\frac{1}{\sqrt{2}}\left( (u_1+iu_2)\ket{00}+(u_4+iu_3)\ket{01}-(u_4-iu_3)\ket{10}+(u_1-iu_2)\ket{11}\right)
\end{equation}
with real numbers $\{u_i\in\rr\}_{i=1}^4$. On the other hand, any maximally entangled state can be obtained by applying the single-qubit unitary operator represented by $
\left(\begin{matrix}
e^{i\phi_1}\cos\theta&&e^{i\phi_2}\sin\theta\\
-e^{-i\phi_2}\sin\theta&&e^{-i\phi_1}\cos\theta
\end{matrix}\right)
$ to $\ket{\Psi_1}$ and can be represented by a vector
\begin{equation}
\label{eq:2QMES}
 \frac{1}{\sqrt{2}}\left(e^{i\phi_1}\cos\theta\ket{00}+e^{i\phi_2}\sin\theta\ket{01}-e^{-i\phi_2}\sin\theta\ket{10}+e^{-i\phi_1}\cos\theta\ket{11}\right).
\end{equation}
By comparing Eqs.~\eqref{eq:rsubspace} and \eqref{eq:2QMES}, we can verify that any two-qubit maximally entangled state can be represented as a unit vector in $\kk_{MES}$ and any unit vector in $\kk_{MES}$ represents a maximally entangled state. This equivalence has been indicated in a previous study \cite{BDSW96}, and the basis defined in Eq.~\eqref{eq:MESbasis} is called the {\it magic basis} \cite{HW97}.

Since $H=\sum_{x\in X}r_x\Phi_x$ is represented as a real symmetric matrix with respect to the basis $\{\ket{\Psi_i}\}_{i=1}^4$, $H$ is diagonalizable with respect to real eigenvectors, which represents maximally entangled states.
\end{proof}

Next, we show a special property of the diamond norm between probabilistic mixtures of single-qubit unitaries in the following Lemma, which essentially shows that the input state in the definition of the diamond norm can be maximally entangled.
\begin{lemma}
\label{lemma:1Qdiamondnorm}
For a subset $\left\{\Upsilon_x:\linop{\cdim{2}}\rightarrow\linop{\cdim{2}}\right\}_{x\in X}$ of single-qubit unitary transformations and probability distributions $p$ and $q$ over a finite set $X$, it holds that
\begin{equation}
\diamondnorm{\sum_{x\in X}p(x)\Upsilon_x-\sum_{x\in X}q(x)\Upsilon_x}=\trdist{\sum_{x\in X}(p(x)-q(x))J(\Upsilon_x)}.
\end{equation}
\end{lemma}
\begin{proof}
 For $d(\geq2)$-dimensional CPTP maps $\{\Upsilon_x\}_{x\in X}$, it holds that
 \begin{equation}
(L.H.S.)=\max_{\Phi\in\puredop{\cd\otimes\cd}}2\trdist{\sum_{x\in X}(p(x)-q(x))\Upsilon_x\otimes id_{\cd}(\Phi)}
\geq\frac{2}{d}\trdist{\sum_{x\in X}(p(x)-q(x))J(\Upsilon_x)}.
\end{equation}

On the other hand, by using the dual problem of the SDP to compute the diamond norm used in the proof of Proposition \ref{prop:SDP}, we obtain
\begin{equation}
(L.H.S.)\leq2\lpnorm{\infty}{\ptr{2}{S}}\ {\rm with}\ \left(S\geq0\right)\wedge \left(S\geq\sum_{x\in X}(p(x)-q(x))J(\Upsilon_x)\right),
\end{equation}
where $\ptr{2}{\cdot}$ represents the partial trace of the second system of $\cdim{2}\otimes\cdim{2}$. By using Lemma \ref{lemma:1QMES}, we can verify that $\sum_{x\in X}(p(x)-q(x))J(\Upsilon_x)=\sum_{i=1}^4\lambda_i\Phi_i$ with real numbers $\lambda_i$ and a set of orthogonal maximally entangled states $\{\Phi_i\}_{i=1}^4$. By setting $S=\sum_{i:\lambda_i>0}\lambda_i\Phi_i$, we obtain
\begin{equation}
 2\lpnorm{\infty}{\ptr{2}{S}}=2\lpnorm{\infty}{\sum_{i:\lambda_i>0}\lambda_i\frac{\idop}{2}}=\sum_{i:\lambda_i>0}\lambda_i=(R.H.S.).
\end{equation}
This completes the proof.
\end{proof}

\subsection{Support of optimal probability distribution}
To achieve the quadratic improvement owing to the probabilistic approximation of $\Upsilon$ by using $\{\Upsilon_x\}_{x\in X}$, we assume $\{\Upsilon_x\}_{x\in X}$ is an $\epsilon$-covering of the set of unitary transformations in Lemma \ref{lemma:upperbound}. Since $|X|=\Omega\left(\frac{1}{\epsilon^c}\right)$ from a volume consideration, the runtime $\poly{|X|\log\left(\frac{1}{\epsilon}\right)}$ of our SDP to compute the optimal probability distribution proposed in Proposition \ref{prop:SDP} increases as $\poly{\frac{1}{\epsilon}}$ at best.
However, by using the following lemma, we can construct a much more efficient SDP.

\begin{figure}[h]
\includegraphics[height=.25\textheight]{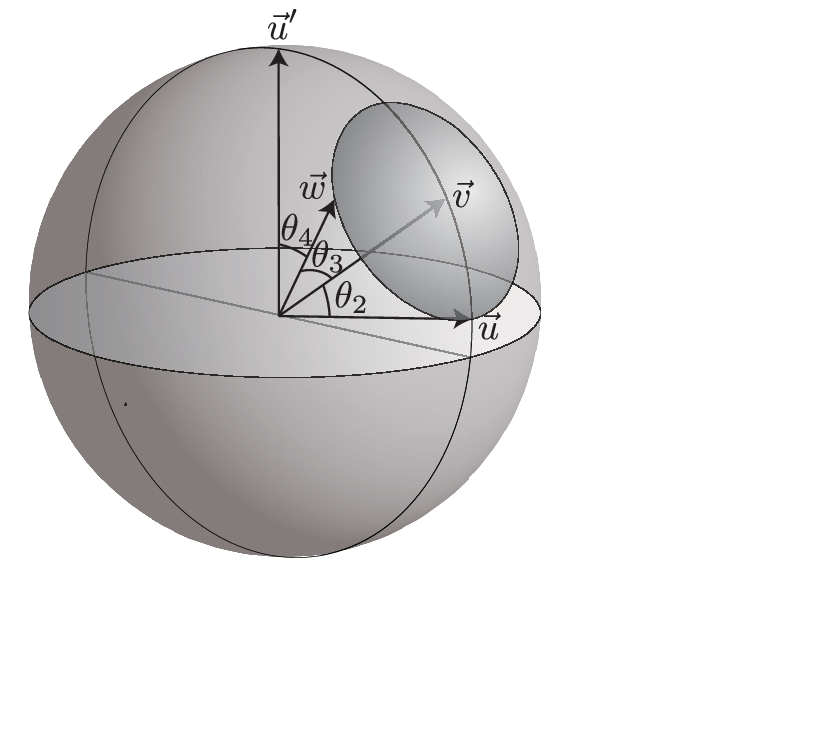}
\caption{\label{fig:1Qsupport}
Three-dimensional subspace spanned by $\{\vec{u},\vec{u}',\vec{w},\vec{v}\}$ in the proof of Lemma \ref{lemma:support}, where $\vec{v}\in\vspan{\{\vec{u},\vec{u}'\}}$. We apply the triangle inequality for the angle $\theta_4$ between $\vec{u}'$ and $\vec{w}$, the angle $\theta_3$ between $\vec{v}$ and $\vec{w}$ and the angle $(\theta_1-\theta_2)$ between $\vec{v}$ and $\vec{u}'$.}
\end{figure}

\begin{lemma}
\label{lemma:support}
For a non-negative real number $\epsilon\geq0$, if $\Upsilon$ is a single-qubit unitary transformation and $\left\{\Upsilon_x:\linop{\cdim{2}}\rightarrow\linop{\cdim{2}}\right\}_{x\in X}$ is a finite $\epsilon$-covering of the set of single-qubit unitary transformations, i.e., $\max_\Upsilon\min_{x\in X}\frac{1}{2}\diamondnorm{\Upsilon-\Upsilon_x}\leq\epsilon$, then
\begin{equation}
\label{eq:support}
\min_{p}\diamondnorm{\Upsilon-\sum_{x\in X}p(x)\Upsilon_x}=\min_{\hat{p}}\diamondnorm{\Upsilon-\sum_{x\in \hat{X}}\hat{p}(x)\Upsilon_x}
\end{equation}
holds, where $\hat{X}:=\{x\in X:\frac{1}{2}\diamondnorm{\Upsilon-\Upsilon_x}\leq2\epsilon\}$ and the minimization of $p$ and $\hat{p}$ are taken over probability distributions over $X$ and those over $\hat{X}$, respectively.
\end{lemma}
\begin{proof}
 By using Lemma \ref{lemma:1Qdiamondnorm}, we obtain
\begin{equation}
\label{eq:6_1}
 (L.H.S.)=\min_{p}\trdist{J(\Upsilon)-\sum_{x\in X}p(x)J(\Upsilon_x)}=\min_{p}\lpnorm{\infty}{J(\Upsilon)-\sum_{x\in X}p(x)J(\Upsilon_x)},
\end{equation}
where we use the dimension of the eigenspace of $J(\Upsilon)-\sum_{x\in X}p(x)J(\Upsilon_x)$ with positive eigenvalues is at most $1$ in the last equality.
By using Lemma \ref{lemma:1QMES}, we can proceed with the following two ways:
\begin{eqnarray}
 {\rm Eq}.~\eqref{eq:6_1}&=&\min_{p}\max_{\rho\in\conv{{\rm MES}}}\tr{\rho \left(J(\Upsilon)-\sum_{x\in X}p(x)J(\Upsilon_x)\right)}\ {\rm and}\\
  {\rm Eq}.~\eqref{eq:6_1}&=&\min_{p}\max_{\substack{M\in{\rm cone}({\rm MES})\\ M\leq \idop}}\tr{M\left(J(\Upsilon)-\sum_{x\in X}p(x)J(\Upsilon_x)\right)},
\end{eqnarray}
where $\conv{{\rm MES}}$ and ${\rm cone}({\rm MES})$ are the convex hull of the set of maximally entangled states $\left\{\Phi\in\puredop{\cdim{2}\otimes\cdim{2}}:\ptr{2}{\Phi}=\frac{\idop}{2}\right\}$ and the convex cone generated by the set $\{\Phi\}$, respectively. Note that the convex cone generated by a subset $\mathbf{X}$ in a vector space is defined as the set of finite linear combinations of $\mathbf{X}$ with non-negative coefficients.

Since the domains of $p$, $\rho$, and $M$ are compact and convex and $f(p,H):=\tr{H(J(\Upsilon)-\sum_{x\in X}p(x)J(\Upsilon_x))}$ is affine with respect to each variable, we can apply the minimax theorem and obtain
\begin{equation}
\label{eq:6_2}
 {\rm Eq}.~\eqref{eq:6_1}=\max_{\rho\in\conv{{\rm MES}}}\left(\tr{\rho J(\Upsilon)}-\max_{x\in X}\tr{\rho J(\Upsilon_x)}\right)
=\max_{\substack{M\in{\rm cone}({\rm MES})\\ M\leq \idop}}\left(\tr{M J(\Upsilon)}-\max_{x\in X}\tr{M J(\Upsilon_x)}\right).
\end{equation}
When $(L.H.S.)=0$, the theorem holds since there exists $x\in X$ such that $\Upsilon_x=\Upsilon$. In the following, we assume $(L.H.S.)>0$. If $\rho$ with $\lpnorm{\infty}{\rho}<1$ maximizes the formula, we can show a contradiction by setting $M=\frac{\rho}{\lpnorm{\infty}{\rho}}$. Thus, $\rho$ that maximizes the formula satisfies $\lpnorm{\infty}{\rho}=1$, i.e., $\rho$ is a (pure) maximally entangled state. Therefore, we obtain
\begin{equation}
\label{eq:6_3}
 {\rm Eq}.~\eqref{eq:6_2}=\max_{\Upsilon'}\frac{1}{2}\left(\tr{J(\Upsilon') J(\Upsilon)}-\max_{x\in X}\tr{J(\Upsilon') J(\Upsilon_x)}\right)=\max_{U'\in U(2)}\frac{1}{2}\left(\left|\tr{U^\dag U'}\right|^2-\max_{x\in X}\left|\tr{U_x^\dag U'}\right|^2\right),
\end{equation}
where $\Upsilon(\rho)=U\rho U^\dag$, $\Upsilon_x(\rho)=U_x\rho U_x^\dag$, the maximization of $\Upsilon'$ is taken over single-qubit unitary transformations, and $U(2)$ represents the set of single-qubit unitary operators. By observing that the minimization in Eq.~\eqref{eq:L0} is achieved by $\rho=\frac{\idop}{2}$ for single-qubit unitary operators, we obtain
\begin{equation}
\label{eq:6_4}
  {\rm Eq}.~\eqref{eq:6_3}=\max_{\Upsilon'}\frac{1}{2}\left(\min_{x\in X}\diamondnorm{\Upsilon'-\Upsilon_x}^2-\diamondnorm{\Upsilon'-\Upsilon}^2\right).
\end{equation}
Since so far we did not use the assumption that $\{\Upsilon_x\}_{x\in X}$ is an $\epsilon$-covering, we obtain
\begin{equation}
 \label{eq:6_5}
 (R.H.S.)\ {\rm of}\  {\rm Eq}.~\eqref{eq:support}=\max_{\Upsilon'}\frac{1}{2}\left(\min_{x\in \hat{X}}\diamondnorm{\Upsilon'-\Upsilon_x}^2-\diamondnorm{\Upsilon'-\Upsilon}^2\right).
\end{equation}
Note that the maximization in Eq.~\eqref{eq:6_4} is achieved by $\Upsilon'$ satisfying $\frac{1}{2}\diamondnorm{\Upsilon'-\Upsilon}\leq\epsilon$ since $\min_{x\in X}\frac{1}{2}\diamondnorm{\Upsilon'-\Upsilon_x}\leq\epsilon$ due to the definition of the $\epsilon$-covering. If we can show that the maximization in Eq.~\eqref{eq:6_5} is also achieved by such $\Upsilon'$, we can prove the equivalence between Eqs.~\eqref{eq:6_4} and \eqref{eq:6_5}. For the minimization in Eq.~\eqref{eq:6_4} is achieved by $x\in \hat{X}$ owing to the triangle inequality. To complete the proof, we show the following statement: for all $\Upsilon'$,
\begin{equation}
\label{eq:obj1}
 \frac{1}{2}\diamondnorm{\Upsilon'-\Upsilon}>\epsilon\Rightarrow \min_{x\in \hat{X}}\diamondnorm{\Upsilon'-\Upsilon_x}\leq\diamondnorm{\Upsilon'-\Upsilon}.
\end{equation}
We assume $\epsilon<1$; otherwise, the statement is trivial.
By using the equivalence between the set of two-qubit maximally entangled vectors and a real subspace shown in the proof of Lemma \ref{lemma:1QMES}, there exist unit real vectors $\vec{u},\vec{u}'\in\rr^4$ such that $\sum_{i,j=1}^4u_iu_j\ket{\Psi_i}\bra{\Psi_j}=\frac{1}{2}J(\Upsilon)$, $\sum_{i,j=1}^4u'_iu'_j\ket{\Psi_i}\bra{\Psi_j}=\frac{1}{2}J(\Upsilon')$ and
\begin{equation}
 0\leq \cos\theta_1:=\vec{u}\cdot \vec{u}'<\sqrt{1-\epsilon^2},
\end{equation}
where $\{\ket{\Psi_i}\}$ is defined in Eq.~\eqref{eq:MESbasis}, $\theta_1\in[0,\frac{\pi}{2}]$, the first inequality can be satisfied by appropriately setting the sign of $\vec{u}$, and the second (strict) inequality is derived from $\frac{1}{2}\diamondnorm{\Upsilon'-\Upsilon}>\epsilon$ and Lemma \ref{lemma:1Qdiamondnorm}. In the real subspace spanned by $\{\vec{u},\vec{u}'\}$, there exists a unique unit real vector $\vec{v}\in\rr^4$ such that
\begin{equation}
 \cos\theta_2:=\vec{u}\cdot \vec{v}=\sqrt{1-\epsilon^2}\ \wedge\ \vec{u}'\cdot \vec{v}=\cos(\theta_1-\theta_2),
\end{equation}
where $\theta_2\in[0,\frac{\pi}{2}]$,
as shown in Fig.~\ref{fig:1Qsupport}. Note that the unitary transformation $\hat{\Upsilon}$ corresponding to $\vec{v}$, i.e., $\sum_{i,j=1}^4v_iv_j\ket{\Psi_i}\bra{\Psi_j}=\frac{1}{2}J(\hat{\Upsilon})$, satisfies  $\frac{1}{2}\diamondnorm{\Upsilon-\hat{\Upsilon}}=\epsilon$ due to Lemma \ref{lemma:1Qdiamondnorm}. Since there exists $x\in X$ such that $\frac{1}{2}\diamondnorm{\Upsilon_x-\hat{\Upsilon}}\leq\epsilon$ and $\frac{1}{2}\diamondnorm{\Upsilon_x-\Upsilon}\leq\frac{1}{2}\diamondnorm{\Upsilon_x-\hat{\Upsilon}}+\frac{1}{2}\diamondnorm{\Upsilon-\hat{\Upsilon}}\leq2\epsilon$, we can find a unit real vector $\vec{w}\in\rr^4$ corresponding to $\Upsilon_x$ with $x\in\hat{X}$, i.e., $\sum_{i,j=1}^4w_iw_j\ket{\Psi_i}\bra{\Psi_j}=\frac{1}{2}J(\Upsilon_x)$, and satisfying
\begin{equation}
\cos\theta_3:= \vec{w}\cdot \vec{v}\geq\sqrt{1-\epsilon^2},
\end{equation}
where $\theta_3\in[0,\frac{\pi}{2}]$,
due to Lemma \ref{lemma:1Qdiamondnorm}. By using Lemma \ref{lemma:1Qdiamondnorm} again, we obtain
\begin{equation}
 \diamondnorm{\Upsilon'-\Upsilon_x}\leq\diamondnorm{\Upsilon'-\Upsilon}\Leftrightarrow|\vec{u}'\cdot \vec{u}|\leq|\vec{u}'\cdot \vec{w}|.
\end{equation}
By letting $\cos\theta_4:=\vec{u}'\cdot \vec{w}$ with $\theta_4\in[0,\pi]$ and using the triangle inequality for angles in the three-dimensional subspace spanned by $\{\vec{u},\vec{u}',\vec{w}\}$, we obtain
\begin{equation}
 \theta_4\leq(\theta_1-\theta_2)+\theta_3\leq\theta_1.
\end{equation}
This completes the proof.

\end{proof}

As an application of Lemma \ref{lemma:support}, we construct an efficient probabilistic synthesis algorithm in the proof of the following theorem.
\begin{theorem}
\label{thm:main2}
For a given gate set, there exists a probabilistic synthesis algorithm for a single-qubit unitary transformation with

{\rm INPUT}: a single-qubit unitary transformation $\Upsilon:\linop{\cdim{2}}\rightarrow\linop{\cdim{2}}$, an approximation error $\epsilon\in\left(0,1\right)$, and precision $\delta>0$ such that $\frac{1}{\delta}=\left(\frac{1}{\epsilon}\right)^{O(1)}$

{\rm OUTPUT}: a gate sequence for implementing a single-qubit unitary transformation $\Upsilon_x$ sampled from a set $\left\{\Upsilon_x:\linop{\cdim{2}}\rightarrow\linop{\cdim{2}}\right\}_{x\in\hat{X}}$ in accordance with probability distribution $\hat{p}(x)$

\noindent
such that the algorithm satisfies the following properties:
\begin{itemize}
 \item {\rm Efficiency}: All steps of the algorithm take $\polylog{\frac{1}{\epsilon}}$-time,
 
 \item {\rm Quadratic improvement}: The approximation error $\frac{1}{2}\diamondnorm{\Upsilon-\sum_{x\in\hat{X}}\hat{p}(x)\Upsilon_x}$ obtained by this algorithm is upper bounded by $\epsilon^2+\delta$, whereas the error $\min_{x\in\hat{X}}\frac{1}{2}\diamondnorm{\Upsilon-\Upsilon_x}$ obtained by deterministic synthesis using the unitary transformations in $\{\Upsilon_x\}_{x\in\hat{X}}$ is upper bounded by $\epsilon$,
 
\end{itemize}
\end{theorem}
\begin{proof}
We assume that the algorithm calls an efficient deterministic synthesis algorithm such as the Solovay-Kitaev algorithm as a subroutine, i.e., the subroutine can find a gate sequence for implementing a unitary transformation $\Upsilon'$ such that $\frac{1}{2}\diamondnorm{\Upsilon-\Upsilon'}\leq\epsilon$ within $\polylog{\frac{1}{\epsilon}}$-time.
In the following, we explicitly construct the algorithm:

\noindent \textbf{Efficient probabilistic synthesis algorithm for single-qubit unitary transformation}
\begin{enumerate}
\item Set free parameters $c>0$ and $c'>0$ satisfying $c+c'\leq1$.

 \item Generate a list $\{\hat{\Upsilon}_x\}_{x\in\hat{X}}$ of single-qubit unitary transformations such that for any unitary transformation $\hat{\Upsilon}$, $\min_{x\in\hat{X}}\frac{1}{2}\diamondnorm{\hat{\Upsilon}-\hat{\Upsilon}_x}\leq c\epsilon$ if $\frac{1}{2}\diamondnorm{\Upsilon-\hat{\Upsilon}}\leq 2\epsilon$. That is, $\{\hat{\Upsilon}_x\}_{x\in\hat{X}}$ is a $c\epsilon$-covering of the $2\epsilon$-ball around the target unitary transformation.
 
 \item Call an efficient deterministic synthesis algorithm to find gate sequences for implementing unitary transformations $\{\Upsilon_x\}_{x\in\hat{X}}$ such that $\frac{1}{2}\diamondnorm{\Upsilon_x-\hat{\Upsilon}_x}\leq c'\epsilon$ for all $x\in\hat{X}$.
 
 \item Numerically solve our SDP shown in Proposition \ref{prop:SDP} by using $\{\Upsilon_x\}_{x\in\hat{X}}$ as a set of CPTP mappings and obtain a probability distribution $\hat{p}$, which causes the approximation error $\delta$-close to $\min_p\frac{1}{2}\diamondnorm{\Upsilon-\sum_{x\in\hat{X}}p(x)\Upsilon_x}$.
 
 \item Sample gate sequences for implementing unitary transformations $\{\Upsilon_x\}_{x\in\hat{X}}$ in accordance with $\hat{p}$.
\end{enumerate}
The two properties can be verified as follows:
\begin{itemize}
 \item {\it Efficiency}: All steps of the algorithm take $\polylog{\frac{1}{\epsilon}}$-time if the size $\hat{X}$ of the list  generated in the second step is upper bounded by a constant (independent to $\epsilon$.) We can generate such a constant-size list $\{\hat{\Upsilon}_x\}_{x\in\hat{X}}$ by using the correspondence between a single-qubit unitary operator and unit vector in $\rdim{4}$ and Lemma \ref{lemma:1Qdiamondnorm}.
 
 \item {\it Quadratic improvement}: The approximation error $\frac{1}{2}\diamondnorm{\Upsilon-\sum_{x\in\hat{X}}\hat{p}(x)\Upsilon_x}$ obtained by this algorithm is at least $\epsilon^2+\delta$ since $\{\Upsilon_x\}_{x\in\hat{X}}$ is a subset of an $\epsilon$-covering $\{\Upsilon_x\}_{x\in\hat{X}}\cup\{\Upsilon'_y\}_y$ of the set of single-qubit unitary transformations, where $\{\Upsilon'_y\}_y$ is an $\epsilon$-covering of the complement of the $2\epsilon$-ball around $\Upsilon$ and $\frac{1}{2}\diamondnorm{\Upsilon-\Upsilon'_y}> 2\epsilon$ for any $y$, and we can apply Lemmas \ref{lemma:upperbound} and \ref{lemma:support}.
 
\end{itemize}

\end{proof}

Note that the quadratic improvement on the approximation error achieved by this algorithm heavily relies on Lemma \ref{lemma:support}. In Appendix \ref{appendix:num}, we perform numerical experiments to confirm that this lemma would hold for qudit unitary transformations, which implies that this synthesis algorithm is applicable to qudit unitary transformations.

\subsection{Convex-hull approximation for axial rotations}
\label{sec:example1Q}
At a glance, the reduction of the approximation error due to probabilistically mixing unitaries seems strange since a unitary transformation is not a probabilistic mixture of any distinct unitary transformations. A simple geometric interpretation of the reduction is given in the following theorem, considering single-qubit unitary transformations corresponding to axial rotations.

\begin{figure}[h]
\includegraphics[height=.2\textheight]{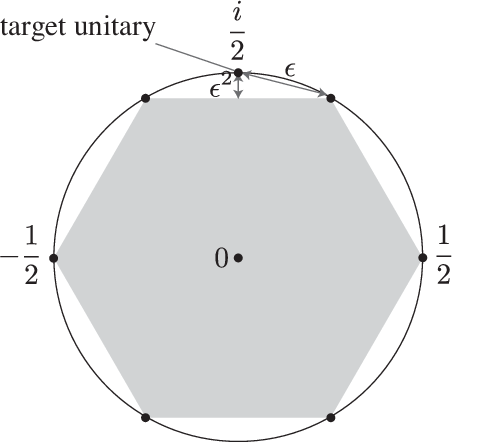}
\caption{\label{fig:singlequbit}
Corresponding complex numbers to target unitary transformation $\Upsilon_{\frac{\pi}{2}}$ and unitary transformations $\{\Upsilon_{\theta}\}_{\theta\in\mathbf{\Theta}}$ with $\mathbf{\Theta}=\left\{0,\frac{\pi}{3},\frac{2\pi}{3},\pi,\frac{4\pi}{3},\frac{5\pi}{3}\right\}$. Convex hull of $\{\Upsilon_{\theta}\}_{\theta\in\mathbf{\Theta}}$ corresponds to shaded region. If we let approximation error obtained by deterministic approximation be $\epsilon=\min_{\theta\in\mathbf{\Theta}}\frac{1}{2}\diamondnorm{\Upsilon_{\frac{\pi}{2}}-\Upsilon_\theta}=\min_{\theta\in\mathbf{\Theta}}\frac{1}{2}\left|i-e^{i\theta}\right|$, that obtained by probabilistic approximation is given by $\epsilon^2=\min_p\frac{1}{2}\diamondnorm{\Upsilon_{\frac{\pi}{2}}-\sum_{\theta\in\mathbf{\Theta}}p(\theta)\Upsilon_\theta}=\min_p\frac{1}{2}\left|i-\sum_{\theta\in\mathbf{\Theta}}p(\theta)e^{i\theta}\right|$, which demonstrates quadratic reduction in error.
}
\end{figure}

We investigate the convex-hull approximation of a single-qubit unitary transformation $\Upsilon_{\hat{\theta}}$ by using unitary transformations $\{\Upsilon_\theta\}_{\theta\in\mathbf{\Theta}}$ that rotate Bloch vectors about the same axes as $\Upsilon_{\hat{\theta}}$, where $\Upsilon_{\theta}(\rho):=R(\theta)\rho R^\dag(\theta)$, $R(\theta):=\ketbra{0}+e^{i\theta}\ketbra{1}$ with an orthonormal basis $\{\ket{0},\ket{1}\}$, and $\mathbf{\Theta}$ is a finite subset of $[0,2\pi)$.
In this case, every unitary transformation $\Upsilon_\theta$ can be represented by a unit complex number $e^{i\theta}$ in the complex plane, as shown in Fig.~\ref{fig:singlequbit}.
Furthermore, the following proposition shows that the metric space of probabilistic mixtures of $\Upsilon_\theta$ induced by the diamond norm can be identified with a unit disc in the complex plane.

\begin{proposition}
\label{prop:axialU}
For a finite subset $\mathbf{\Theta}$ of $[0,2\pi)$, let $\{\Upsilon_\theta\}_{\theta\in \mathbf{\Theta}}$ be a set of single-qubit unitary transformations that rotate Bloch vectors about a fixed axis, i.e., $\Upsilon_{\theta}(\rho):=R(\theta)\rho R^\dag(\theta)$ with $R(\theta):=\ketbra{0}+e^{i\theta}\ketbra{1}$ and an orthonormal basis $\{\ket{0},\ket{1}\}$. For probability distributions $p$ and $q$ over $\mathbf{\Theta}$, it holds that
\begin{equation}
\diamondnorm{\sum_{\theta\in\mathbf{\Theta}}p(\theta)\Upsilon_\theta-\sum_{\theta\in\mathbf{\Theta}}q(\theta)\Upsilon_\theta}=\left|\sum_{\theta\in\mathbf{\Theta}}p(\theta)e^{i\theta}-\sum_{\theta\in\mathbf{\Theta}}q(\theta)e^{i\theta}\right|.
\end{equation}
\end{proposition}
\begin{proof}
 By using Lemma \ref{lemma:1Qdiamondnorm}, we obtain
 \begin{equation}
 (L.H.S.)=\trdist{\sum_{\theta\in\mathbf{\Theta}}(p(\theta)-q(\theta))J(\Upsilon_\theta)}=(R.H.S.),
\end{equation}
where we use the diagonalization of $\sum_{\theta\in\mathbf{\Theta}}(p(\theta)-q(\theta))J(\Upsilon_\theta)$, which can be obtained via a straightforward calculation, in the last equality.
\end{proof}

By using this proposition, we can obtain $\frac{1}{2}\diamondnorm{\Upsilon_{\hat{\theta}}-\sum_{\theta\in\mathbf{\Theta}}p(\theta)\Upsilon_\theta}=\frac{1}{2}\left|e^{i\hat{\theta}}-\sum_{\theta\in\mathbf{\Theta}}p(\theta)e^{i\theta}\right|$, which indicates that the optimal probability distribution and approximation error in the convex-hull approximation of $\Upsilon_{\hat{\theta}}$ can be computed by finding the closest point in the convex hull of $\{e^{i\theta}\}_{\theta\in\mathbf{\Theta}}$ to the target point $e^{i\hat{\theta}}$. As represented in Fig.~\ref{fig:singlequbit}, the quadratic reduction in  approximation error owing to convex-hull approximation over discrete-point approximation can be shown by an elementary geometric observation.

\section{Conclusion}
\label{sec:conclusion}
We considered the analytical relationship between $\min_p\diamondnorm{\Upsilon-\sum_{x}p(x)\Upsilon_x}$ and $\min_x\diamondnorm{\Upsilon-\Upsilon_x}$, which represent the minimum approximation error obtained by probabilistic synthesis and that by deterministic synthesis, respectively. As the main result, we obtained tight upper and lower bounds on $\min_p\diamondnorm{\Upsilon-\sum_{x}p(x)\Upsilon_x}$, which guarantees the sub-optimality of the current algorithms as well as suggests the existence of an improved synthesis algorithm.
We showed that the optimal probability distribution in the approximation can be computed by an SDP. We also constructed an efficient probabilistic synthesis algorithm for single-qubit unitary transformations and showed that it quadratically reduces approximation error compared with deterministic synthesis and its optimality can be reduced into the choice of unitary transformations close to the target unitary one. While numerical simulations indicate the algorithm works well for qudit unitary transformations, a rigorous proof is a subject for future work.

When we run this algorithm for qudit unitary transformation, the time complexity of the SDP used in the algorithm becomes $\poly{d,|\hat{X}|}$ for a fixed desired approximation error $\epsilon$ and precision $\delta$, where $d$ is the dimension of the unitary operators and $|\hat{X}|$ is the size of the list of synthesized unitary transformations.
Since $|\hat{X}|$ grows exponentially with respect to $d^2$ (to make the list an $\epsilon$-covering of the $2\epsilon$-ball around the target unitary transformation), the algorithm is not practical for higher dimensional unitary transformations.

There are two ways to make the algorithm more practical. First, restricting a class of target unitary transformations, such as axial rotations, would significantly reduce the size $|\hat{X}|$ but still achieve the guaranteed quadratic improvement. Indeed, Fig.~\ref{fig:singlequbit} implies that the quadratic improvement can be achieved by mixing only two realizable unitary transformations. Alternatively, we can consider a modified algorithm that probabilistically mixes a randomly sampled small subset of $\hat{X}$. While this modified algorithm does not provide the guaranteed quadratic improvement as the original one, numerical experiments in Appendix \ref{appendix:num} suggest that it still attains such improvement for randomly chosen target unitary transformations.

Similar to the probabilistic mixture of unitary transformations, that of general CPTP mappings implemented by a certain quantum device is relatively easy to implement by classically controlling the quantum device. Such a probabilistic mixture of implementable CPTP mappings is considered a {\it free operation} in many quantum resource theories \cite{HO13, BG15, CG19}. To quantify or simulate a target CPTP mapping using the probabilistic mixture (sometimes assisted by a resource state), a mathematical tool is required to analyze the optimal convex approximation of a general CPTP mapping. From the mathematical perspective as well as from the resource theoretical perspective, computing or bounding the approximation error of a {\it unital} CPTP mapping by using a probabilistic mixture of unitary transformations plays a crucial role in investigating the asymptotic quantum Birkhoff conjecture \cite{HM11, NDQ12}.
Our SDP shown in Proposition \ref{prop:SDP} and our bounds (or possibly their extension to general CPTP mappings) could be numerical and analytical tools to investigate such problems.

\begin{acks}
We thank Yoshihisa Yamamoto, Aram Harrow, Isaac Chuang, Sho Sugiura, Yuki Takeuchi, Yasunari Suzuki, Yasuhiro Takahashi, and Adel Sohbi for their helpful discussions.
This work was partially supported by JST Moonshot R\&D MILLENNIA Program (Grant No.JPMJMS2061).
SA was partially supported by JST, PRESTO Grant No.JPMJPR2111 and JPMXS0120319794.
GK was supported in part by the Grant-in-Aid for Scientific Research (C) No.20K03779, (C) No.21K03388, and (S) No.18H05237 of JSPS, CREST (Japan Science and Technology Agency) Grant No.JPMJCR1671.
ST was partially supported by JSPS KAKENHI Grant Numbers JP20H05966 and JP22H00522.
\end{acks}

\bibliographystyle{ACM-Reference-Format}
\bibliography{references.bib}

\appendix

\section{Equivalence between quantum testers and quantum networks}
\label{appendix:tester}
Recall that  the Choi-Jamio\l kowski operator of linear mapping $\Xi:\linop{\hh_1}\rightarrow\linop{\hh_2}$ is defined as $J(\Xi):=\sum_{i,j}\ket{i}\bra{j}\otimes\Xi(\ket{i}\bra{j})\in\linop{\hh_1\otimes\hh_2}$, and the set of quantum testers is defined as $ \mathbf{T}(\hh_1:\hh_2):=\{T\in\pos{\hh_1\otimes\hh_2}:\exists\rho\in\dop{\hh_1},T\leq\rho\otimes\idop_{\hh_2}\}$.
 In this section, we show that the set of mappings $f_T:\Xi\mapsto\tr{J\left(\Xi\right)T}$ associated with quantum testers $T\in \mathbf{T}(\hh_1:\hh_2)$ is equivalent to that of mappings $g_{\Phi,\Pi}:\Xi\mapsto\tr{\Xi\otimes id_{\hh_3}(\Phi)\Pi}$ associated with pure states $\Phi\in\puredop{\hh_1\otimes\hh_3}$ and Hermitian projectors $\Pi\in\mathbf{Proj}(\hh_2\otimes\hh_3)$ for sufficiently large dimensional Hilbert space $\hh_3$. This equivalence indicates
 \begin{equation}
 \label{eq:tester_and_network}
 \max_{T\in \mathbf{T}(\hh_1:\hh_2)}\min_{\Xi}f_T(\Xi)=\max_{\substack{\Phi\in\puredop{\hh_1\otimes\hh_3}\\\Pi\in\mathbf{Proj}(\hh_2\otimes\hh_3)}}\min_{\Xi}g_{\Phi,\Pi}(\Xi),
\end{equation}
where the minimization of $\Xi$ is taken over a compact subset of linear mappings specified in the proofs of Proposition \ref{prop:SDP} and Lemma \ref{lemma:upperbound}.
Note that a proof for more general quantum testers is given in \cite[Theorem~10]{GGP09}.

First, we show that for any $\Phi$ and $\Pi$, there exists $T\in \mathbf{T}(\hh_1:\hh_2)$ such that $f_T=g_{\Phi,\Pi}$ as follows.
By letting $T=\ptr{3}{(\Phi^{T_1}\otimes\idop_2)(\idop_1\otimes\Pi)}$, we obtain
\begin{eqnarray}
g_{\Phi,\Pi}(\Xi)=\tr{\Xi\otimes id_{\hh_3}(\Phi)\Pi}&=&\tr{(J(\Xi)\otimes\idop_3)(\Phi^{T_1}\otimes\idop_2)(\idop_1\otimes\Pi)}=\tr{J(\Xi)T}=f_T(\Xi),
\end{eqnarray}
where $\Phi^{T_1}$ and $\ptr{3}{\cdot}$ represent the partial transpose of $\Phi$ and the partial trace, respectively, and the subscript of the operator denotes the system on which the operator acts. We can also verify that $T\in \mathbf{T}(\hh_1:\hh_2)$ as follows. Let $X=\sum_{ij}\alpha_{ij}\ket{j}_3\bra{i}_1$, where $\ket{\Phi}=\sum_{ij}\alpha_{ij}\ket{i}_1\ket{j}_3$ with the computational basis $\{\ket{i}_1\in\puredop{\hh_1}\}_i$ and $\{\ket{j}_3\in\puredop{\hh_3}\}_j$. We then obtain that for any positive semidefinite operator $P\in\pos{\hh_1\otimes\hh_2}$,
\begin{equation}
\tr{PT}=\tr{(P\otimes\idop_3)(\Phi^{T_1}\otimes\idop_2)(\idop_1\otimes\Pi)}=\tr{(X\otimes\idop_2)P(X\otimes\idop_2)^\dag\Pi}\geq0,
\end{equation}
which indicates $T\geq0$. By letting $\rho=\ptr{3}{\Phi^{T_1}}=\ptr{3}{\Phi}^T(\in\dop{\hh_1})$, we can also verify that
\begin{eqnarray}
\rho\otimes\idop_2-T=\ptr{3}{(\Phi^{T_1}\otimes\idop_2)(\idop_{123}-\idop_1\otimes\Pi)}=\ptr{3}{(\Phi^{T_1}\otimes\idop_2)(\idop_1\otimes\Pi_\bot)}\geq0,
\end{eqnarray}
where $\Pi_\bot\in\mathbf{Proj}(\hh_2\otimes\hh_3)$ satisfies $\Pi+\Pi_\bot=\idop$, and the last inequality can be verified by the fact that $T\geq0$.

Next, we show that for any $T\in \mathbf{T}(\hh_1:\hh_2)$, there exist $\Phi\in\puredop{\hh_1\otimes\hh_3}$ and $\Pi\in\mathbf{Proj}(\hh_2\otimes\hh_3)$ such that $f_T=g_{\Phi,\Pi}$ as follows. Let $T\leq\rho_1\otimes\idop_2$, $\hat{\Phi}\in\puredop{\hh_1\otimes\hh_{1'}}$ be a purification of $\rho_1^T$, its singular value decomposition be $\ket{\hat{\Phi}}=\sum_i\sqrt{p(i)}\ket{x_i}_1\ket{y_i}_{1'}$ ($p(i)>0$), and $P\in\pos{\hh_2\otimes\hh_{1'}}$ be $P=XTX^\dag$, where $X=\sum_{i}\frac{1}{\sqrt{p(i)}}\ket{y_i}_{1'}\bra{x_i^*}_1$ and $\ket{\phi^*}$ is the complex conjugate of $\ket{\phi}$. We can then verify that
\begin{equation}
f_T(\Xi)=\tr{J(\Xi)T}=\tr{(J(\Xi)\otimes\idop_{1'})(\hat{\Phi}^{T_1}\otimes\idop_2)(\idop_1\otimes P)}=\tr{\Xi\otimes id_{\hh_{1'}}(\hat{\Phi})P}.
\end{equation}
Since $P\leq X(\rho_1\otimes\idop_2)X^\dag\leq\idop_{1'2}$, $\{P,\idop-P\}$ is a positive operator-valued measure (POVM).
Owing to the Naimark's extension, we can embed $\hat{\Phi}$ and $\{P,\idop-P\}$ in a larger Hilbert space as a pure state $\Phi$ and a projection-valued measure (PVM) $\{\Pi,\Pi_\bot\}$, respectively, which completes the proof.

\section{Formal SDPs and their strong duality}
\label{appendix:SDP}
A formal SDP to compute $\frac{1}{2}\diamondnorm{\mathcal{A}-\mathcal{B}}$ is defined with a triple $(\Xi,A,B)$ such that
\begin{eqnarray}
A=\left(
\begin{matrix}
 J(\mathcal{A}-\mathcal{B})&0&0\\
 0&0&0\\
 0&0&0
\end{matrix}
\right),\ \ 
B=\left(
\begin{matrix}
 0&0\\
 0&1
\end{matrix}
\right),\ \ 
\Xi\left(\left(
\begin{matrix}
 T&*&*\\
 *&T'&*\\
 *&*&\rho
\end{matrix}
\right)\right)=
\left(
\begin{matrix}
 T+T'-\rho\otimes\idop_{\hh_2}&0\\
 0&\tr{\rho}
\end{matrix}
\right)
\end{eqnarray}
holds for any linear operators $T,T'\in\linop{\hh_1\otimes\hh_2}$ and $\rho\in\linop{\hh_1}$, where the asterisks in the argument to $\Xi$ represent arbitrary linear operators upon which $\Xi$ does not depend, and we identify a linear operator and its matrix representation with respect to a fixed orthonormal basis. The dual problem is obtained by observing that the adjoint of $\Xi$ satisfies
\begin{equation}
 \Xi^\dag\left(\left(
\begin{matrix}
 S&*\\
 *&r\\
\end{matrix}
\right)\right)=
\left(
\begin{matrix}
 S&0&0\\
 0&S&0\\
 0&0&r\idop_{\hh_1}-\ptr{\hh_2}{S}
\end{matrix}
\right)
\end{equation}
for any linear operator $S\in\linop{\hh_1\otimes\hh_2}$ and any complex number $r\in\cc$. We can verify the strong duality of this SDP by observing $\Xi\left(\frac{\idop_{\hh_1}\otimes\idop_{\hh_2}}{2\dim\hh_1}\oplus\frac{\idop_{\hh_1}\otimes\idop_{\hh_2}}{2\dim\hh_1}\oplus\frac{\idop_{\hh_1}}{\dim\hh_1}\right)=B$ and applying the Slater's theorem.

A formal SDP shown in Proposition \ref{prop:SDP} is defined with a triple $(\Xi,A,B)$ such that
\begin{eqnarray}
A&=&\left(
\begin{matrix}
 J(\mathcal{A})&0&0&0&0\\
 0&0&0&0&0\\
 0&0&0&0&0\\
 0&0&0&0&0\\
 0&0&0&0&-1
\end{matrix}
\right)\\
B&=&\left(
\begin{matrix}
 0&0&0\\
 0&1&0\\
 0&0&0
\end{matrix}
\right)\\
  \Xi^\dag\left(\left(
\begin{matrix}
 S&*&*\\
 *&r&*\\
 *&*&P
\end{matrix}
\right)\right)&=&
\left(
\begin{matrix}
 S+\sum_{x\in X}P(x)J(\mathcal{B}_x)&0&0&0&0\\
 0&S&0&0&0\\
 0&0&r\idop_{\hh_1}-\ptr{\hh_2}{S}&0&0\\
 0&0&0&P&0\\
 0&0&0&0&-\tr{P}
\end{matrix}
\right)
\end{eqnarray}
holds for any linear operators $S\in\linop{\hh_1\otimes\hh_2}$, $P\in\linop{\cdim{|X|}}$ and any complex number $r\in\cc$,
where $P(x)$ represents a diagonal element $\bra{x}P\ket{x}$. The primal problem is obtained by observing that the adjoint of $\Xi^\dag$ satisfies
\begin{equation}
 \Xi\left(
 \left(
\begin{matrix}
 T&*&*&*&*\\
 *&T'&*&*&*\\
 *&*&\rho&*&*\\
 *&*&*&Q&*\\
 *&*&*&*&t
\end{matrix}
\right)
 \right)=
 \left(
\begin{matrix}
 T+T'-\rho\otimes\idop_{\hh_2}&0&0\\
 0&\tr{\rho}&0\\
 0&0&\sum_{x\in X}\tr{J(\mathcal{B}_x)T}\ketbra{x}+Q-t\idop_{\cdim{|X|}}
\end{matrix}
\right)
\end{equation}
for any linear operators $T,T'\in\linop{\hh_1\otimes\hh_2}$, $\rho\in\linop{\hh_1}$, $Q\in\linop{\cdim{|X|}}$ and any complex number $t\in\cc$.
We can verify the strong duality of this SDP by observing $\Xi\left(\frac{\idop_{\hh_1}\otimes\idop_{\hh_2}}{2\dim\hh_1}\oplus\frac{\idop_{\hh_1}\otimes\idop_{\hh_2}}{2\dim\hh_1}\oplus\frac{\idop_{\hh_1}}{\dim\hh_1}\oplus\frac{\idop_{\cdim{|X|}}}{2}\oplus 1\right)=B$ and applying the Slater's theorem.

\section{Sharpness of approximation error bounds}
\label{appendix:sharp}
In this section, we make the same assumption $d\geq2$ as Lemma \ref{lemma:lowerbound} and \ref{lemma:upperbound}.
\subsection{Lower bounds}
To show the sharpness of the lower bounds in Ineqs.~\eqref{ineq:lowerbound} and \eqref{ineq:worstbound}, 
we consider a set $\{\Upsilon_x\}_{x\in X}:=\{\Upsilon:\exists W\in\mathbf{W}_\epsilon^{(d)},\Upsilon(\rho)=W\rho W^\dag\}$ of unitary transformations, where
\begin{eqnarray}
\mathbf{W}_\epsilon^{(d)}:=\left\{W:
 W\in U(d)\wedge
 \min_{z\in\conv{\lambda(W)}}|z|\leq\sqrt{1-\epsilon^2},
\right\}\ \ {\rm with}\ \ \epsilon\in[0,1]
\ {\rm and}\ d\geq2,
\end{eqnarray}
where $U(d)$ represents the set of unitary operators acting on $\cd$, $\lambda(W)$ represents the set of eigenvalues of $W$, and $\conv{X}$ represents the convex hull of a subset $X$ in a vector space. To be precise, the two lower bounds are not directly applicable to $\{\Upsilon_x\}_{x\in X}$ since the size $|X|$ of the set is infinite. However, the compactness of the set of unitary transformations on a finite-dimensional Hilbert space enables us to extend Ineqs.~\eqref{ineq:lowerbound} and \eqref{ineq:worstbound} for $|X|=\infty$ by replacing $\min_{x\in X}\frac{1}{2}\diamondnorm{\Upsilon-\Upsilon_x}$ and $\min_{p}\frac{1}{2}\diamondnorm{\Upsilon-\sum_{x\in X}p(x)\Upsilon_x}$ with $\inf_{x\in X}\frac{1}{2}\diamondnorm{\Upsilon-\Upsilon_x}$ and $\inf_{\Lambda\in\conv{\{\Upsilon_x\}_{x\in X}}}\frac{1}{2}\diamondnorm{\Upsilon-\Lambda}$, respectively. 

We show that this example achieves the lower bounds in the extended inequalities with a target unitary transformation $\Upsilon=id$ in Ineq.~\eqref{ineq:lowerbound}. This also indicates that there exists a finite subset $\{\Upsilon_x\}_{x\in \tilde{X}}$ of $\{\Upsilon_x\}_{x\in X}$ such that $\min_{p}\frac{1}{2}\diamondnorm{id-\sum_{x\in \tilde{X}}p(x)\Upsilon_x}$ and $\max_\Upsilon\min_{p}\frac{1}{2}\diamondnorm{\Upsilon-\sum_{x\in \tilde{X}}p(x)\Upsilon_x}$ are arbitrarily close to their each lower bound in Ineqs.~\eqref{ineq:lowerbound} and \eqref{ineq:worstbound}, respectively. For letting $\{\Upsilon_x\}_{x\in \tilde{X}}$ be an $\tilde{\epsilon}$-covering of $\{\Upsilon_x\}_{x\in X}$ with sufficiently small $\tilde{\epsilon}$ is sufficient to show this. Thus, the sharpness of the lower bounds in the extended inequalities indicates that in the original inequalities.
Note that we can show the sharpness of Ineq.~\eqref{ineq:lowerbound} when an $\Upsilon$ is not the identity transformation by replacing $\{\Upsilon_x\}_{x\in X}$ with $\{\Upsilon\circ\Upsilon_x\}_{x\in X}$.

First, by using Eq.~\eqref{eq:L0}, we obtain
\begin{eqnarray}
&&\max_{\Upsilon} \inf_{x\in X}\frac{1}{2}\diamondnorm{\Upsilon-\Upsilon_x}\geq
\inf_{x\in X}\frac{1}{2}\diamondnorm{id-\Upsilon_x}\nonumber\\
\label{ineq:A_epsilon}
&&=\sqrt{1-\sup_{W\in\mathbf{W}_\epsilon^{(d)}}\min_{\rho\in\dop{\cd}}\left|\tr{\rho W}\right|^2}
=\sqrt{1-\sup_{W\in\mathbf{W}_\epsilon^{(d)}}\min_{z\in\conv{\lambda(W)}}|z|^2}=\epsilon.
\end{eqnarray}

Second, by using the extended version of Eq.~\eqref{eq:worstaccuracy}, we obtain
\begin{eqnarray}
\label{ineq:A_convex}
\inf_{\Lambda\in\conv{\{\Upsilon_x\}_{x\in X}}}\frac{1}{2}\diamondnorm{id-\Lambda}\leq
\max_\Upsilon\inf_{\Lambda\in\conv{\{\Upsilon_x\}_{x\in X}}}\frac{1}{2}\diamondnorm{\Upsilon-\Lambda}
=1-\min_{V\in U(d),\rho\in\dop{\cd}}\sup_{W\in\mathbf{W}_\epsilon^{(d)}}\left|\tr{\rho V^\dag W}\right|^2.
\end{eqnarray}
In the following, we show that for any $V\in U(d)$ and $\rho\in\dop{\cd}$,
\begin{equation}
\label{ineq:A_supW}
\sup_{W\in\mathbf{W}_\epsilon^{(d)}}\left|\tr{\rho V^\dag W}\right|^2\geq \left(1-\frac{2\delta}{d}\right)^2\ {\rm with}\ \delta=1-\sqrt{1-\epsilon^2},
\end{equation}
which is sufficient to verify that $\{\Upsilon_x\}_{x\in X}$ achieves lower bounds in the extended Ineqs.~\eqref{ineq:lowerbound} and \eqref{ineq:worstbound}.

Let the diagonalization of $V$ be $V=\sum_{i=1}^d\lambda_i(V)\ket{i}\bra{i}$. Since $\sup_{W\in\mathbf{W}_\epsilon^{(d)}}|\tr{\rho V^\dag W}|^2=\tr{\rho V^\dag V}|^2=1$ if $\min_{z\in\conv{\lambda(V)}}|z|\leq\sqrt{1-\epsilon^2}$, we assume $\epsilon>0$ and $\min_{z\in\conv{\lambda(V)}}|z|>\sqrt{1-\epsilon^2}$. We can then define $\{W^{(ij)}\in\mathbf{W}_\epsilon^{(d)}\}_{1\leq i<j\leq d}$ as 
\begin{eqnarray}
W^{(ij)}&:=&\sum_{k\notin \{i,j\}}\lambda_k(V)\ket{k}\bra{k}+\lambda_+^{(ij)}\ket{i}\bra{i}+\lambda_-^{(ij)}\ket{j}\bra{j}, \\
{\rm where}\ 
\lambda_\pm^{(ij)}&=&\sqrt{1-\epsilon^2}\frac{\lambda_i(V)+\lambda_j(V)}{|\lambda_i(V)+\lambda_j(V)|}\pm\epsilon\frac{\lambda_i(V)-\lambda_j(V)}{|\lambda_i(V)-\lambda_j(V)|}
\ {\rm if}\ \lambda_i(V)\neq\lambda_j(V),\\
{\rm and}\  \lambda_\pm^{(ij)}&=&\sqrt{1-\epsilon^2}\lambda_i(V)\pm i\epsilon\lambda_i(V)\ \ \ \ \ \ \ \ \ \ \ \ \ \ \ \ \ \ \ \ \ \ \ \ \ \ \ {\rm if}\ \lambda_i(V)=\lambda_j(V).
\end{eqnarray}

\begin{figure}[h]
\includegraphics[height=.18\textheight]{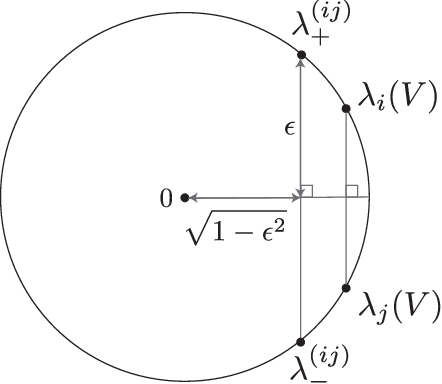}
\caption{\label{fig:eigenvals} Geometric positions of eigenvalues $\lambda_i(V)$, $\lambda_j(V)$ and $\lambda_\pm^{(ij)}$ of unitary operators, which lie on unit circle in complex plane. Note that real and imaginary axes are rotated to horizontalize line equidistant from $\lambda_i(V)$ and $\lambda_j(V)$.}
\end{figure}

(See geometric positions of eigenvalues in the complex plane shown in Fig.~\ref{fig:eigenvals}.)
Note that we can easily verify that $|\lambda_\pm^{(ij)}|=1$ and $\left|\frac{1}{2}\left(\lambda_+^{(ij)}+\lambda_-^{(ij)}\right)\right|=\sqrt{1-\epsilon^2}$, which guarantees $W^{(ij)}\in\mathbf{W}_\epsilon^{(d)}$. Moreover, we can verify that $\lambda\left(V^\dag W^{(ij)}\right)=\{1,z^{(ij)},z^{(ij)*}\}$ with a unit complex number $z^{(ij)}$ satisfying $\re{z^{(ij)}}\geq\sqrt{1-\epsilon^2}$.
Then, for any  $V\in U(d)$ and $\rho\in\dop{\cd}$, the left hand side of Ineq.~\eqref{ineq:A_supW} can be bounded as
\begin{eqnarray}
\sup_{W\in\mathbf{W}_\epsilon^{(d)}}\left|\tr{\rho V^\dag W}\right|^2&\geq&\max_{1\leq i<j\leq d}\left|\tr{\rho V^\dag W^{(ij)}}\right|^2
\geq \min_p\max_{1\leq i<j\leq d}\left|\sum_{k\notin\{i,j\}}p(k)+p(i)z^{(ij)}+p(j)z^{(ij)*}\right|^2\\
&\geq& \min_p\max_{1\leq i<j\leq d}\left\{\sum_{k\notin\{i,j\}}p(k)+(p(i)+p(j))\re{z^{(ij)}}\right\}^2\\
&\geq& \min_p\max_{1\leq i<j\leq d}\left\{\sum_{k\notin\{i,j\}}p(k)+(p(i)+p(j))\sqrt{1-\epsilon^2}\right\}^2\\
&\geq& \min_p\max_{1\leq i<j\leq d}\left\{1-\delta(p(i)+p(j))\right\}^2
\geq \left(1-\frac{2\delta}{d}\right)^2.
\end{eqnarray}
This completes the proof.

\subsection{Upper bound}
We show the sharpness of the upper bound in Ineq.~\eqref{ineq:worstbound},
We consider a set $\{\Upsilon_x\}_{x\in X}:=\{\Upsilon:\exists V\in\mathbf{V}_\epsilon^{(d)},\Upsilon(\rho)=V\rho V^\dag\}$ of unitary transformations, where
\begin{eqnarray}
\mathbf{V}_\epsilon^{(d)}&:=&\left\{
\left(\begin{matrix}
1&0\\
0&V_1
\end{matrix}\right)
\left(\begin{matrix}
W&0\\
0&\idop_{d-2}
\end{matrix}\right)
\left(\begin{matrix}
1&0\\
0&V_2
\end{matrix}\right):
V_1,V_2\in U(d-1),W\in \mathbf{R}_\epsilon
\right\},
\\
\mathbf{R}_\epsilon&:=&\left\{
\left(\begin{matrix}
\cos\theta&-\sin\theta\\
\sin\theta&\cos\theta
\end{matrix}\right)
:0\leq\theta\leq\arccos(\epsilon)\right\}
\ \ {\rm with}\ \ \epsilon\in[0,1]\ {\rm and}\ d\geq2.
\end{eqnarray}
Here $\idop_d$ represents the $d\times d$ identity matrix, and we identify a unitary operator and its matrix representation with respect to a fixed orthonormal basis $\{\ket{i}\}_{i=0}^{d-1}$.
Since $|X|=\infty$, we show the sharpness of the upper bound in the extended Ineq.~\eqref{ineq:worstbound}, which is defined in the proof of the sharpness of the lower bounds.
Note that
\begin{equation}
\label{eq:CU_dec}
\forall U\in U(d),\exists\alpha\in\rr,\exists V\in\mathbf{V}_0^{(d)},U=e^{i\alpha}V
\end{equation}
holds.
This can be verified from the following three observations: First, by letting $U\ket{i}=\ket{e_i}$, there exists $\tilde{V}_1,\tilde{V}_2\in U(d-1)$ and $\tilde{W}\in U(2)$ such that $\left(\begin{matrix}
\tilde{W}^\dag&0\\
0&\idop_{d-2}
\end{matrix}\right)
\left(\begin{matrix}
1&0\\
0&\tilde{V}_1^\dag
\end{matrix}\right)\ket{e_0}=\ket{0}$ and 
$\left(\begin{matrix}
1&0\\
0&\tilde{V}_2^\dag
\end{matrix}\right)
\left(\begin{matrix}
\tilde{W}^\dag&0\\
0&\idop_{d-2}
\end{matrix}\right)
\left(\begin{matrix}
1&0\\
0&\tilde{V}_1^\dag
\end{matrix}\right)\ket{e_i}=\ket{i}$ for all $i$.
Second, for any $\tilde{W}\in U(2)$, there exists $\alpha,\beta,\gamma\in\rr$ and $W\in\mathbf{R}_0$ such that
$\tilde{W}=e^{i\alpha}
\left(\begin{matrix}
1&0\\
0&e^{i\beta}
\end{matrix}\right)
W
\left(\begin{matrix}
1&0\\
0&e^{i\gamma}
\end{matrix}\right)
$.
Third, by letting $V_1=\tilde{V}_1
\left(\begin{matrix}
e^{i\beta}&0\\
0&e^{-i\alpha}\idop_{d-2}
\end{matrix}\right)
$ and $V_2=
\left(\begin{matrix}
e^{i\gamma}&0\\
0&\idop_{d-2}
\end{matrix}\right)
\tilde{V}_2$, we can verify $U=e^{i\alpha}\left(\begin{matrix}
1&0\\
0&V_1
\end{matrix}\right)
\left(\begin{matrix}
W&0\\
0&\idop_{d-2}
\end{matrix}\right)
\left(\begin{matrix}
1&0\\
0&V_2
\end{matrix}\right)$.

First, by using Eq.~\eqref{eq:L0}, we obtain
\begin{eqnarray}
\max_{\Upsilon} \inf_{x\in X}\frac{1}{2}\diamondnorm{\Upsilon-\Upsilon_x}
&=&\sqrt{1-\min_{U\in U(d)}\sup_{V\in\mathbf{V}_\epsilon^{(d)}}\min_{\rho\in\dop{\cd}}\left|\tr{\rho U^\dag V}\right|^2}\\
&=&\sqrt{1-\min_{W\in \mathbf{R}_0}\sup_{V\in\mathbf{V}_\epsilon^{(d)}}\min_{\rho\in\dop{\cd}}\left|\tr{\rho
\left(\begin{matrix}
W^\dag&0\\
0&\idop_{d-2}
\end{matrix}\right)
V}\right|^2}\\
&\leq&\sqrt{1-\min_{W\in \mathbf{R}_0}\sup_{W'\in\mathbf{R}_\epsilon}\min_{\rho\in\dop{\cd}}\left|\tr{\rho
\left(\begin{matrix}
W^\dag W'&0\\
0&\idop_{d-2}
\end{matrix}\right)
}\right|^2}\\
&=&\sqrt{1-\min_{0\leq\theta\leq\frac{\pi}{2}}\sup_{0\leq\theta'\leq\arccos(\epsilon)}\min_{\rho\in\dop{\cd}}\left|\tr{\rho
\left(\begin{matrix}
\cos(\theta'-\theta)&-\sin(\theta'-\theta)&0\\
\sin(\theta'-\theta)&\cos(\theta'-\theta)&0\\
0&0&\idop_{d-2}
\end{matrix}\right)
}\right|^2}\\
&=&\sqrt{1-\min_{0\leq\theta\leq\frac{\pi}{2}}\sup_{0\leq\theta'\leq\arccos(\epsilon)}\cos^2(\theta'-\theta)}\\
&=&\max_{0\leq\theta\leq\frac{\pi}{2}}\inf_{0\leq\theta'\leq\arccos(\epsilon)}|\sin(\theta'-\theta)|=\epsilon,
\end{eqnarray}
where we use Eq.~\eqref{eq:CU_dec} in the second equality and use $\lambda\left(
\left(\begin{matrix}
\cos(\theta'-\theta)&-\sin(\theta'-\theta)&0\\
\sin(\theta'-\theta)&\cos(\theta'-\theta)&0\\
0&0&\idop_{d-2}
\end{matrix}\right)
\right)=\{1,e^{\pm i(\theta'-\theta)}\}$ in the fourth equality.
 
Second, by using the definition of the diamond norm, we obtain
\begin{eqnarray}
\label{ineq:AL_convex}
\max_\Upsilon\inf_{\Lambda\in\conv{\{\Upsilon_x\}_{x\in X}}}\frac{1}{2}\diamondnorm{\Upsilon-\Lambda}
&\geq& \max_\Upsilon\inf_{\Lambda\in\conv{\{\Upsilon_x\}_{x\in X}}}\trdist{\Upsilon(\ketbra{0})-\Lambda(\ketbra{0})}\\
&\geq&1- \min_\Upsilon\sup_{\Lambda\in\conv{\{\Upsilon_x\}_{x\in X}}}\fidelity{\Upsilon(\ketbra{0})}{\Lambda(\ketbra{0})}\\
&=&1- \min_\Upsilon\sup_{x\in X}\fidelity{\Upsilon(\ketbra{0})}{\Upsilon_x(\ketbra{0})}\\
&=&1- \min_{U\in U(d)}\sup_{V\in \mathbf{V}_\epsilon^{(d)}}\left|\bra{0}U^\dag V\ket{0}\right|^2\\
&=&1- \min_{W\in \mathbf{R}_0}\sup_{\substack{W'\in \mathbf{R}_\epsilon\\ V\in U(d-1)}}\left|\bra{0}
\left(\begin{matrix}
W^\dag&0\\
0&\idop_{d-2}
\end{matrix}\right)
\left(\begin{matrix}
1&0\\
0&V
\end{matrix}\right)
\left(\begin{matrix}
W'&0\\
0&\idop_{d-2}
\end{matrix}\right)
 \ket{0}\right|^2\\
 &=&1-\min_{0\leq\theta\leq\frac{\pi}{2}}\sup_{\substack{0\leq\theta'\leq\arccos(\epsilon)\\ V\in U(d-1)}}\left|\cos\theta\cos\theta'+\sin\theta\sin\theta'\bra{1}V\ket{1}\right|^2\\
 &=&\max_{0\leq\theta\leq\frac{\pi}{2}}\inf_{0\leq\theta'\leq\arccos(\epsilon)}\sin^2(\theta'-\theta)=\epsilon^2,
\end{eqnarray}
where we use $\trdist{\phi-\rho}=\max_{\Pi\in\proj{\hh}}\tr{\Pi(\phi-\rho)}\geq1-\tr{\phi\rho}$ in the second inequality and use Eq.~\eqref{eq:CU_dec} in the third equality.
This and the extended upper bound in  Ineq.~\eqref{ineq:worstbound} complete the proof.

\section{Numerical experiment for actual approximation errors}
\label{appendix:num}
Recall that Theorem \ref{thm:main1} has established the relationship between the actual approximation error $\epsilon_\Upsilon^{prob}$ caused by the probabilistic approximation, that $\epsilon_\Upsilon$ caused by the deterministic approximation, and the worst approximation error $\epsilon$, where each approximation error is defined by
\begin{eqnarray}
 \epsilon_\Upsilon^{prob}&:=& \min_{p}\frac{1}{2}\diamondnorm{\Upsilon-\sum_{x\in X}p(x)\Upsilon_x}\\
  \epsilon_\Upsilon&:=&\min_{x\in X}\frac{1}{2}\diamondnorm{\Upsilon-\Upsilon_x}\\
   \epsilon&:=&\max_{\Upsilon}\min_{x\in X}\frac{1}{2}\diamondnorm{\Upsilon-\Upsilon_x}.
\end{eqnarray}
By using these notations, we can rewrite the relationship established in Theorem \ref{thm:main1} as follows:
\begin{equation}
 \frac{2\epsilon_\Upsilon^2}{d}\simeq\frac{4\delta_\Upsilon}{d}\left(1-\frac{\delta_\Upsilon}{d}\right)\leq  \epsilon_\Upsilon^{prob}\leq\epsilon^2,
\end{equation}
where $\delta_\Upsilon=1-\sqrt{1-\epsilon_\Upsilon^2}$ and the approximation becomes tighter when $\epsilon_\Upsilon\rightarrow 0$.

While these inequalities are tight, it is helpful to know how small $\epsilon_\Upsilon^{prob}$ can be realized compared to $\epsilon_\Upsilon^2$. In this appendix, we perform numerical experiments to demonstrate that $\epsilon_\Upsilon^{prob}$ is comparable to $\epsilon_\Upsilon^2$ for randomly chosen target unitary transformations $\Upsilon$. 
Moreover, we demonstrate that $\epsilon_\Upsilon^{prob}$ becomes smaller than $\epsilon_\Upsilon^2$ for high-dimensional unitary transformations, i.e., the probabilistic approximation reduces the approximation error more than quadratically.

Additionally, the experiments have another purpose: to provide pieces of evidence supporting that Lemma \ref{lemma:support} holds for qudit unitary transformations, which is crucial in constructing our probabilistic synthesis algorithm. Recall that it states that
\begin{equation}
 \min_{p}\diamondnorm{\Upsilon-\sum_{x\in X}p(x)\Upsilon_x}=\min_{\hat{p}}\diamondnorm{\Upsilon-\sum_{x\in \hat{X}}\hat{p}(x)\Upsilon_x},
\end{equation}
where $\hat{X}:=\{x\in X:\frac{1}{2}\diamondnorm{\Upsilon-\Upsilon_x}\leq2\epsilon\}$.
Our numerical experiments support this statement for randomly sampled target unitary transformations $\Upsilon:\linop{\cd}\rightarrow\linop{\cd}$, randomly constructed $\epsilon$-coverings $\left\{\Upsilon_x:\linop{\cd}\rightarrow\linop{\cd}\right\}_{x\in X}$, and $d=3,4$.

\noindent\textbf{Setting of numerical experiments}

First, we construct $\epsilon$-coverings $\left\{\Upsilon_x:\linop{\cd}\rightarrow\linop{\cd}\right\}_{x\in X}$ by randomly choosing $|X|=10^5$, $|X|=10^6$ and $|X|=10^7$ unitary operators for $d=2$, $d=3$, and $d=4$, respectively. Note that the random sampling of unitary operators means the sampling probability distribution is induced by the Haar measure on $U(d)$.
We compute a lower bound on $\epsilon$ as $\max_i\min_{x\in X}\frac{1}{2}\diamondnorm{\Upsilon_i-\Upsilon_x}$ by using $30$ randomly chosen target unitary transformations $\{\Upsilon_i\}_{i=1}^{30}$. 
We interpret $\left\{\Upsilon_x\right\}_{x\in X}$ as the set of available unitary transformations in probabilistic and deterministic approximation.

Next, we randomly choose a target unitary transformation $\Upsilon$ and compute $\epsilon_\Upsilon$.
We define the actual approximation error caused by probabilistically mixing restricted available unitary transformations as
\begin{equation}
  \epsilon_\Upsilon^{prob}(\epsilon'):= \min_{p}\frac{1}{2}\diamondnorm{\Upsilon-\sum_{x\in \hat{X}(\epsilon')}p(x)\Upsilon_x}
\end{equation}
where $\hat{X}(\epsilon'):=\{x\in X:\frac{1}{2}\diamondnorm{\Upsilon-\Upsilon_x}\leq\epsilon'\}$.
Note that $\epsilon_\Upsilon^{prob}(\epsilon')$ is a monotonically decreasing function. Moreover, if Lemma \ref{lemma:support} holds for qudit unitary transformations, $\epsilon_\Upsilon^{prob}(2\epsilon)=\epsilon_\Upsilon^{prob}(1)=\epsilon_\Upsilon^{prob}$.

Third, we compute the actual approximation error caused by probabilistically mixing more restricted available unitary transformations as
\begin{equation}
  \epsilon_\Upsilon^{prob}(N):= \min_{p}\frac{1}{2}\diamondnorm{\Upsilon-\sum_{x=1}^Np(x)\Upsilon_x},
\end{equation}
where $\{\Upsilon_x\}_{x=1}^N$ is a randomly sampled subset of $\hat{X}(\epsilon')$ and $\epsilon'$ is chosen large enough for $\epsilon_\Upsilon^{prob}(\epsilon')$ to converge.

\noindent\textbf{Results of numerical experiments}

\begin{figure}[h]
\includegraphics[height=.65\textheight]{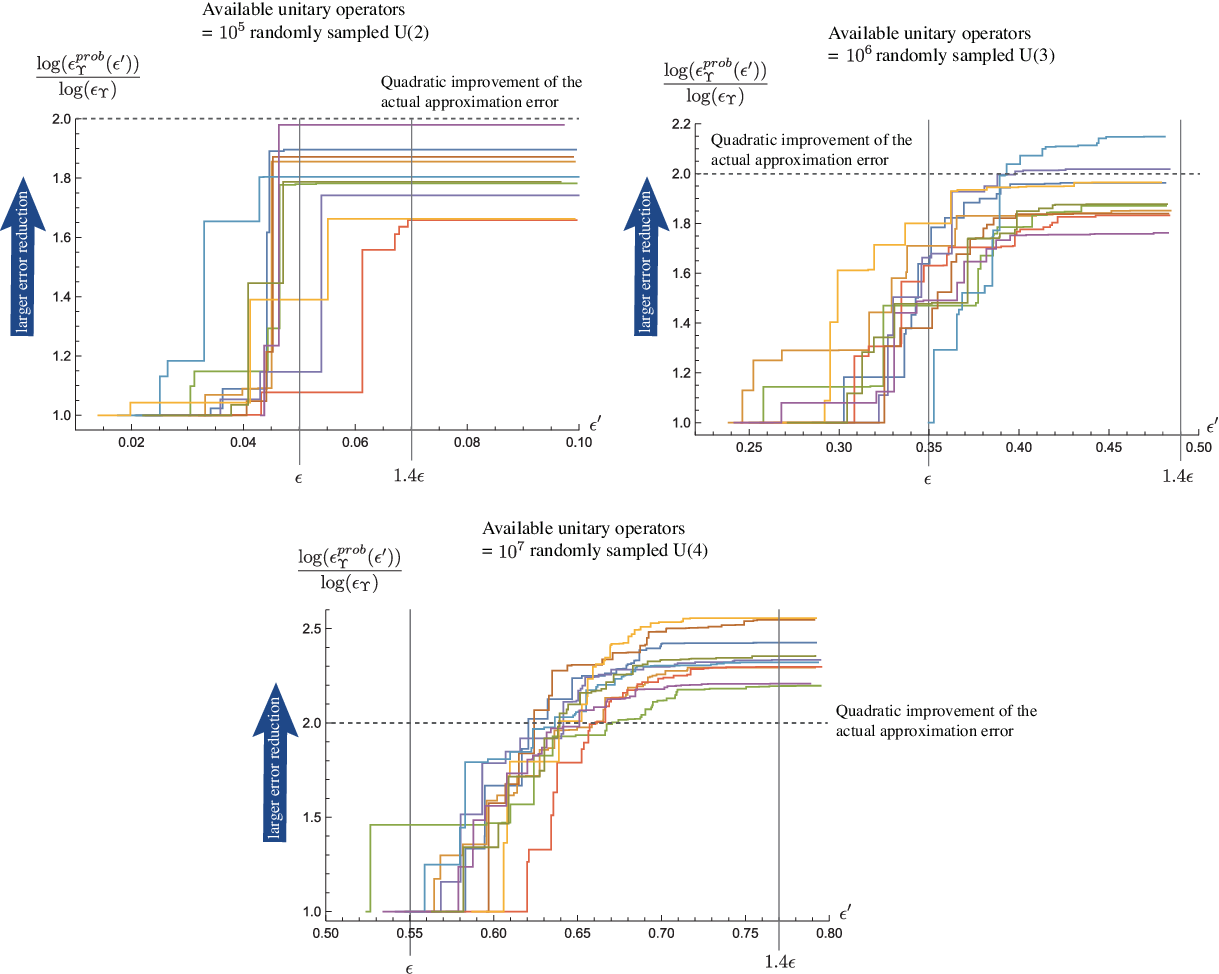}
\caption{\label{fig:synth} The comparison between the actual approximation error $\epsilon_\Upsilon^{prob}$ caused by the probabilistic approximation and that $\epsilon_\Upsilon$ caused by the deterministic approximation for $10$ randomly sampled target unitary transformations $\Upsilon$. For both approximations, we use the set of available unitary transformations induced by an $\epsilon$-covering of $U(d)$.
For each $\Upsilon$, we compute the actual approximation errors $\epsilon_\Upsilon^{prob}(\epsilon')$ caused by probabilistically mixing the set of available unitary transformations $\Upsilon_x$ such that $\frac{1}{2}\diamondnorm{\Upsilon-\Upsilon_x}\leq\epsilon'$. The gray vertical lines represent lower bounds on $\epsilon$ and $1.4\epsilon$.}
\end{figure}

In Fig.~\ref{fig:synth}, we draw the graphs of $\epsilon_\Upsilon^{prob}(\epsilon')$ for $10$ randomly chosen $\Upsilon$ by using different colors corresponding to $\Upsilon$. We can observe that $\epsilon_\Upsilon^{prob}(\epsilon')$ is saturated when $\epsilon'\geq1.4\epsilon$ and $\epsilon_\Upsilon^{prob}$ is comparable or smaller than $\epsilon_\Upsilon^2$ since $\frac{\log(\epsilon_\Upsilon^{prob}(\epsilon'))}{\log(\epsilon_\Upsilon)}\geq2\Leftrightarrow\epsilon_\Upsilon^{prob}(\epsilon')\leq \epsilon_\Upsilon^2$ and $\epsilon_\Upsilon^{prob}\leq \epsilon_\Upsilon^{prob}(\epsilon')$ by definition.

\begin{figure}[h]
\includegraphics[height=.4\textheight]{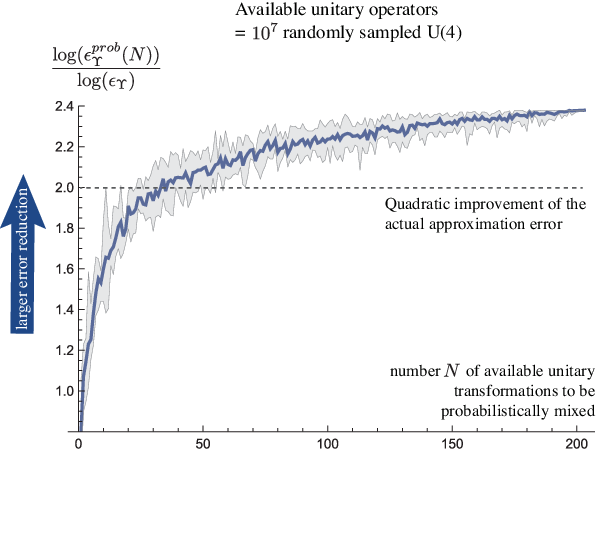}
\caption{\label{fig:4dimsynth} The approximation error caused by probabilistically mixing restricted subsets $\{\Upsilon_x\}_{x=1}^N$ of available unitary transformations. First, we randomly choose a target unitary transformation $\Upsilon$. Next, we randomly choose 8 such subsets from the set $\{\Upsilon_x:\frac{1}{2}\diamondnorm{\Upsilon-\Upsilon_x}\leq\epsilon'\}$ of available unitary transformations that seem sufficient to attain the optimal approximation error caused by probabilistically mixing all the available unitary transformations. The blue graph represents the average of the approximation error and the gray region represents its variance.}
\end{figure}

In Fig.~\ref{fig:4dimsynth}, we draw the graph of $\epsilon_\Upsilon^{prob}(N)$ for a randomly chosen $\Upsilon:\linop{\cdim{4}}\rightarrow \linop{\cdim{4}}$. Note that we plot its empirical variance and average since $\epsilon_\Upsilon^{prob}(N)$ is a random variable depending on the choice of a subset of $\hat{X}(\epsilon')$.
We can observe that the approximation error $\epsilon_\Upsilon^{prob}(N)$ rapidly converges to its minimum $\epsilon_\Upsilon^{prob}(\epsilon')$.
Therefore, we can expect that our probabilistic synthesis algorithm proposed in Theorem \ref{thm:main2} can be made more efficient while still attaining an approximation error that is nearly optimal.

\noindent\textbf{Environment of numerical experiments}

We performed numerical experiments using Mathematica 14.0.0.0 on a MacBook Pro equipped with a 2.4GHz Intel Core i9 processor and 64GB of memory.

\end{document}